\documentclass[twocolumn,showpacs,preprintnumbers,amsmath,amssymb]{revtex4}

\usepackage{graphicx}
\usepackage{dcolumn}
\usepackage{bm}
\usepackage{slashbox}

\usepackage{makeidx}
\makeindex
\usepackage{graphics}

\def\Vec#1{\mbox{\boldmath $#1$}}

\begin{document}
\preprint{OU-TAP 229}
\preprint{YITP-04-15}
\title{Coincidence analysis to search for inspiraling compact binaries
using TAMA300 and LISM data}
\author{Hirotaka Takahashi$^{1,2,3}$\footnote{hirotaka@vega.ess.sci.osaka-u.ac.jp}}
\author{Hideyuki Tagoshi$^{1}$}

\author{Masaki Ando$^{4}$}
\author{Koji Arai$^{5}$}
\author{Peter Beyersdorf$^{5}$}
\author{Nobuyuki Kanda$^{6}$}
\author{Seiji Kawamura$^{5}$}
\author{Norikatsu Mio$^{7}$}
\author{Shinji Miyoki$^{8}$}
\author{Shigenori Moriwaki$^{7}$}
\author{Kenji Numata$^{4}$}
\author{Masatake Ohashi$^{8}$}
\author{Misao Sasaki$^{3}$}
\author{Shuichi Sato$^{5}$}
\author{Ryutaro Takahashi$^{5}$}
\author{Daisuke Tatsumi$^{5}$}
\author{Yoshiki Tsunesada$^{5}$}

\author{Akito Araya$^{9}$}
\author{Hideki Asada$^{10}$}
\author{Youich Aso$^{4}$}
\author{Mark A. Barton$^{8}$}
\author{Masa-Katsu Fujimoto$^{5}$}
\author{Mitsuhiro Fukushima$^{5}$}
\author{Toshifumi Futamase$^{11}$}
\author{Tomiyoshi Haruyama$^{12}$}
\author{Kazuhiro Hayama$^{13}$}
\author{Gerhard Heinzel$^{5}$\footnote{Currently at 
Max-Planck-Institut f\"ur Gravitationsphysik (Albert-Einstein-Inst.)
Institut Hannover, Am kleinen Felde 30, D-30167 Hannover, Germany}}
\author{Gen'ichi Horikoshi$^{12}$\footnote{deceased}}
\author{Yukiyoshi Iida$^{4}$}
\author{Kunihito Ioka$^{1}$}
\author{Hideki Ishitsuka$^{8}$}
\author{Norihiko Kamikubota$^{12}$}
\author{Kunihiko Kasahara$^{8}$}
\author{Keita Kawabe$^{4}$}
\author{Nobuki Kawashima$^{14}$}
\author{Yasufumi Kojima$^{15}$}
\author{Kazuhiro Kondo$^{8}$}
\author{Yoshihide Kozai$^{5}$}
\author{Kazuaki Kuroda$^{8}$}
\author{Namio Matsuda$^{16}$}
\author{Kazuyuki Miura$^{17}$}
\author{Osamu Miyakawa$^{8}$\footnote{Currently at Department of Physics, California Institute of Technology, Pasadena, CA 91125, USA}}
\author{Shoken Miyama$^{5}$}
\author{Mitsuru Musha$^{18}$}
\author{Shigeo Nagano$^{5}$}
\author{Ken'ichi Nakagawa$^{18}$}
\author{Takashi Nakamura$^{19}$}
\author{Hiroyuki Nakano$^{6}$}
\author{Ken-ichi Nakao$^{6}$}
\author{Yuhiko Nishi$^{4}$}
\author{Yujiro Ogawa$^{12}$}
\author{Naoko Ohishi$^{5}$}
\author{Akira Okutomi$^{8}$}
\author{Ken-ichi Oohara$^{20}$}
\author{Shigemi Otsuka$^{4}$}
\author{Yoshio Saito$^{12}$}
\author{Nobuaki Sato$^{12}$}
\author{Hidetsugu Seki$^{4}$}
\author{Naoki Seto$^{1}$}
\author{Masaru Shibata$^{21}$}
\author{Takakazu Shintomi$^{12}$}
\author{Kenji Soida$^{4}$}
\author{Kentaro Somiya$^{7}$}
\author{Toshikazu Suzuki$^{12}$}
\author{Akiteru Takamori$^{4}$}
\author{Shuzo Takemoto$^{19}$}
\author{Kohei Takeno$^{7}$}
\author{Takahiro Tanaka$^{19}$}
\author{Toru Tanji$^{7}$}
\author{Shinsuke Taniguchi$^{4}$}
\author{Colin T. Taylor$^{8}$}
\author{Souichi Telada$^{22}$}
\author{Kuniharu Tochikubo$^{4}$}
\author{Takayuki Tomaru$^{12}$}
\author{Yoji Totsuka$^{12}$}
\author{Kimio Tsubono$^{4}$}
\author{Nobuhiro Tsuda$^{23}$}
\author{Takashi Uchiyama$^{8}$}
\author{Akitoshi Ueda$^{5}$}
\author{Ken-ichi Ueda$^{18}$}
\author{Fumihiko Usui$^{21}$}
\author{Koichi Waseda$^{5}$}
\author{Yuko Watanabe$^{17}$}
\author{Hiromi Yakura$^{17}$}
\author{Kazuhiro Yamamoto$^{8}$}
\author{Akira Yamamoto$^{12}$}
\author{Toshitaka Yamazaki$^{5}$}
\author{Tatsuo Yoda$^{4}$}
\author{Zong-Hong Zhu$^{5}$}

\affiliation{
${}^1$Department of Earth and Space Science, Graduate School of Science, 
Osaka University, Toyonaka, Osaka 560-0043, Japan}
\affiliation{
${}^2$Graduate School of Science and Technology,
Niigata University, Niigata, Niigata 950-2181, Japan}
\affiliation{
${}^{3}$Yukawa Institute for Theoretical Physics, Kyoto University, Kyoto, Kyoto 606-8502, Japan}
\affiliation{
${}^{4}$Department of Physics, University of Tokyo, Hongo, Bunkyo-ku, Tokyo 113-0033, Japan}
\affiliation{
${}^{5}$National Astronomical Observatory of Japan, Mitaka, Tokyo 181-8588, Japan}
\affiliation{
${}^{6}$Department of Physics,  Graduate School of Science, Osaka City University, Sumiyoshi-ku, Osaka, Osaka 558-8585, Japan}
\affiliation{
${}^{7}$Department of Advanced Materials Science, University of Tokyo,  Kashiwa, Chiba 277-8561, Japan}
\affiliation{
${}^{8}$Institute for Cosmic Ray Research, University of Tokyo, Kashiwa, Chiba 277-8582, Japan}
\affiliation{
${}^{9}$Earthquake Research Institute, University of Tokyo, Bunkyo-ku, Tokyo 113-0032, Japan}
\affiliation{
${}^{10}$Faculty of Science and Technology, Hirosaki University, Hirosaki, Aomori 036-8561, Japan}
\affiliation{
${}^{11}$Astronomical Institute, Tohoku University, Sendai, Miyagi 980-8578, Japan}
\affiliation{
${}^{12}$High Energy Accelerator Research Organization, Tsukuba, Ibaraki 305-0801, Japan}
\affiliation{
${}^{13}$Department of Astronomy, University of Tokyo, Bunkyo-ku, Tokyo 113-0033, Japan}
\affiliation{
${}^{14}$Department of Physics, Kinki University, Higashi-Osaka, Osaka 577-8502, Japan}
\affiliation{
${}^{15}$Department of Physics, Hiroshima University, Higashi-Hiroshima, Hiroshima 739-8526, Japan}
\affiliation{
${}^{16}$Department of Materials Science and Engineering, Tokyo Denki University, Chiyoda-ku, Tokyo 101-8457, Japan}
\affiliation{
${}^{17}$ Department of Physics, Miyagi University of Education, Aoba Aramaki, Sendai 980-0845, Japan}
\affiliation{
${}^{18}$Institute for Laser Science, University of Electro-Communications, Chofugaoka, Chofu, Tokyo 182-8585, Japan}
\affiliation{
${}^{19}$Department of Physics, Kyoto University, Kyoto, Kyoto 606-8502, Japan}
\affiliation{
${}^{20}$Department of Physics, Niigata University, Niigata, Niigata 950-2102, Japan}
\affiliation{
${}^{21}$Graduate School of Arts and Sciences, University of Tokyo, Komaba, Meguro, Tokyo 153-8902, Japan}
\affiliation{
${}^{22}$National Institute of Advanced Industrial Science and Technology, Tsukuba , Ibaraki 305-8563, Japan}
\affiliation{
${}^{23}$Precision Engineering Division, Tokai University, Hiratsuka, Kanagawa 259-1292, Japan
}

\author{(The TAMA collaboration and the LISM collaboration)}

\date{\today}

\hspace{0.5cm}

\begin{abstract}
Japanese laser interferometric gravitational wave detectors, 
TAMA300 and LISM, performed a coincident observation during 2001. 
We perform a coincidence analysis to search for inspiraling compact binaries. 
The length of data used for the coincidence analysis is 275 hours
when both TAMA300 and LISM detectors are operated simultaneously. 
TAMA300 and LISM data 
are analyzed by matched filtering, and candidates for gravitational wave 
events are obtained. 
If there is a true gravitational wave signal, it should appear 
in both data of detectors 
with consistent waveforms characterized by masses of stars, 
amplitude of the signal, the coalescence time and so on.
We introduce a set of coincidence conditions of the parameters,
and search for coincident events.
This procedure reduces the number of fake events considerably, 
by a factor $\sim 10^{-4}$ compared with the number 
of fake events in single detector analysis. 
We find that the number of events after imposing 
the coincidence conditions is consistent 
with the number of accidental coincidences produced purely by noise. 
We thus find no evidence of gravitational wave signals. 
We obtain an upper limit of 0.046 /hours (CL $= 90 \%$) 
to the Galactic event rate within 1kpc from the Earth. 
The method used in this paper 
can be applied straightforwardly to the case of coincidence
observations with more than two detectors with arbitrary arm directions.
\end{abstract}

\pacs{95.85.Sz, 04.80.Nn, 07.05.Kf, 95.55.Ym}

\maketitle

\widetext

\section{Introduction}

In the past several years, there has been substantial progress
in gravitational wave detection experiments by the
ground-based laser interferometers,
LIGO\cite{ref:LIGOdescription}, VIRGO\cite{ref:virgo}, 
GEO600\cite{ref:geo},
and TAMA300\cite{ref:tama, ref:ando}. 
The observation of gravitational waves will 
not only be a powerful tool to test general relativity,
but also be a new tool
to investigate various unsolved astronomical problems 
and to find new objects which were not seen
 by other observational methods. 

The Japanese two laser interferometers, TAMA300 and LISM, 
performed a coincident 
observation during August 1 and September 20, 2001 (JST).
Both detectors showed sufficient stability that was acceptable 
for an analysis to search for gravitational wave signals.
Given the sufficient amount of data, it was a very good opportunity to 
perform a coincidence analysis with real interferometers' data. 

There were several works to search for gravitational waves 
using interferometeric data. 
A coincidence analysis searching for generic gravitational wave bursts
in a pair of laser interferometers has been reported in \cite{ref:Nicholson}.
Allen et al.\cite{ref:40m} analyzed LIGO 40m data 
and obtained an upper limit of 0.5/hour (CL = 90$\%$)
on the Galactic event rate of the coalescence of neutron star binaries
with mass between 1$M_\odot$ and 3$M_\odot$. 
Tagoshi et al.\cite{ref:DT2} analyzed TAMA300 data taken during 1999 and 
obtained an upper limit of 0.59/hour (CL = 90$\%$) 
on the event rate of inspirals of compact 
binaries with mass between 0.3$M_\odot$ and 10$M_\odot$ and 
with signal-to-noise ratio greater than 7.2. 
Very recently, an analysis using the first scientific data of
the three LIGO detectors was 
reported~\cite{ref:LIGOinspiral}, and an upper limit of 
$1.7\times 10^2$ per year per Milky Way Equivalent Galaxy is reported. 
Recently, International Gravitational Event Collaboration (IGEC)
of bar detectors reported their analysis using four years of 
data to search for gravitational wave bursts \cite{ref:IGEC03}. 
They found that the event rate they obtained was consistent
with the background of the detectors' noise. 


In the matched filtering analysis using real data 
of single laser interferometer (e.g. \cite{ref:40m}, \cite{ref:DT2}),
many fake events were produced by non-Gaussian and non-stationary noise. 
In order to remove such fake events, 
it is useful to perform coincidence analysis between 
two or more independent detectors. 
Furthermore, coincidence analysis is indispensable to confirm the detection 
of gravitational waves when candidates for real gravitational wave 
signals are obtained. 
The purpose of this paper is to perform coincidence analysis using 
the real data of TAMA300 and LISM. 

We consider gravitational waves from inspiraling compact binaries,
comprized of neutron stars or black holes. They are consider to be one of the 
most promising sources for ground based laser interferometers. 
Since the waveforms of the inspiraling compact binaries are known accurately, 
we employ the matched filtering by using the theoretical waveforms as templates.
Matched filtering is the optimal detection strategy 
in the case of stationary and Gaussian noise of detector.
However, since the detectors' noise is not stationary and Gaussian in the real
laser interferomters, we introduce $\chi^2$ selection method to the matched filtering.

We analyze the data from each detector by matched filtering
which produces event lists. 
Each event is characterized by the time of coalescence, 
masses of the two stars, and the amplitude of the signal. 
If there is a real gravitational wave event,
there must be an event in each of the event list with 
consistent values of parameters. 
We define a set of coincidence conditions to search for coincident events in the
two detectors. 
We find that we
can reduce the number of events to about $10^{-4}$ times the original 
number.
The coincidence conditions are tested by 
injecting the simulated inspiraling waves into the data 
and by checking the detection efficiency. 
We find that the detection efficiency is not affected significantly
by imposing the coincidence conditions. 

We estimate the number of coincident events 
produced accidentally by the instrumental noise.
By using a technique of shifting the time series of data artificially,
we find that the number of events survived after imposing   
the coincidence conditions is consistent 
with the number of accidental coincidences produced purely by noise. 

We propose a method to set an upper limit to the 
real event rate using results of the coincidence analysis. 
In the case of TAMA300 and LISM, we obtain an upper limit of the event rate 
as 0.046/hour (CL = 90$\%$) for inspiraling compact binaries with mass
between 1$M_\odot$ and 2$M_\odot$ which are located within 1kpc from
the Earth. 
In this case, since TAMA300 is much more sensitive than LISM, 
the upper limit obtained from the coincidence analysis is less 
stringent than that obtained from the TAMA300 single detector data analysis.
This is because the detection efficiency in the coincidence analysis
is determined by the sensitivity of LISM.
Thus, the upper limit obtained here is not the optimal one
which we could obtain using the TAMA300 data taken during 2001. 

The method to set an upper limit to the event rate 
proposed here can be extended
straightforwardly to the case of a coincidence analysis
for a network of interferometric gravitational wave detectors. 

This paper is organized as follows. 
In Section~\ref{sec:detector}, 
we briefly describe the TAMA300 and LISM detectors. 
In Section~\ref{sec:method},
we discuss a method of 
matched filtering search used for TAMA300 and LISM data.
In Section~\ref{sec:matchedfilterresults}, 
the results of the matched filtering search for each detector
are shown. In section~\ref{sec:coincidence}, 
we discuss a method of the coincidence analysis using the results 
of single-detector searches, and the result of 
the coincidence analysis is shown. 
We also derive the upper limit to the event rate in Section~\ref{sec:limit}.
Section~\ref{sec:summary} is devoted to summary.
In Appendix A, we discuss a $\chi^2$ veto method to distinguish between
real events and fake events produced by non-Gaussian noise. 
In Appendix B, we examine a different choice of $\Delta t$
( the length of duration to find local maximum of matched filtering output )
 for comparison.
In Appendix C, we discuss a sidereal time distribution of coincidence events.
In Appendix D, we review a method to estimate 
the errors in the parameters due to noise using the Fisher matrix. 

Throughout this paper, the Fourier transform of a function $h(t)$ is denoted
by $\tilde{h} (f)$, which is defined by 
\begin{equation}
\tilde{h} (f) = \int_{-\infty}^{\infty} dt\ e^{2 \pi i f t} h(t) .
\end{equation}

\section{Detector}
\label{sec:detector}

\subsection{TAMA300}\label{sec:detectort}

TAMA300 is a Fabry-Perot-Michelson 
interferometer with the baseline length of $300{\rm m}$ located 
at the National Astronomical Observatory of Japan in Mitaka, Tokyo
($35.68^\circ$N, $139.54^\circ$E) (See Table \ref{tab:summary}). 
The detector's arm orientation 
(the direction of the bisector of two arms) 
measured counterclockwise from East is $225^\circ$.
The details of TAMA300 detector configuration can be found in \cite{ref:ando}.
The TAMA300 detector became ready to operate in the summer
 1999 \cite{ref:tama}. 
Most of the designed system (except power recycling) were installed by the
that time. First data taking was performed as a test during August 1999 (DT1). 
In September 1999, three days observation (DT2) was carried out, 
and the first search for gravitational waves from inspiraling
compact binaries was performed \cite{ref:DT2}. 
Since then, TAMA300 has been performing several observations. 
In August 2000, an observation (DT4) was performed for two weeks 
and 160 hours of data were taken which are described in detail in 
\cite{ref:ando}. From March 2nd to March 8th, 2001, 
TAMA300 performed an observation (DT5) and 111 hours of data 
were taken.
After improvements of the sensitivity, TAMA300 had carried out a long 
observation (DT6) from August 1st to September 20th, 2001.
The length of data taken was about 1100 hours.
The best strain equivalent sensitivity 
was about $h\sim 5\times10^{-21}/\sqrt{\rm Hz}$ around $800$Hz at DT6. 
{}From August 31th to September 2nd, 2002, 
TAMA300 performed a short observation (DT7) and  24 hours of data were taken. 
{}From February 14th and April 15th 2003, TAMA300 performed
 an observation
(DT8) for two months, and 1158 hours of data were taken. 
Most recently, from November 28th 2003 to 10th January, 2004,  
TAMA300 performed an observation (DT9) and 557 hours of data 
were taken.
The observation history of TAMA300 is summarized
in Table \ref{tab:history}. 

\begin{figure}
\scalebox{0.5}[0.5]{\includegraphics{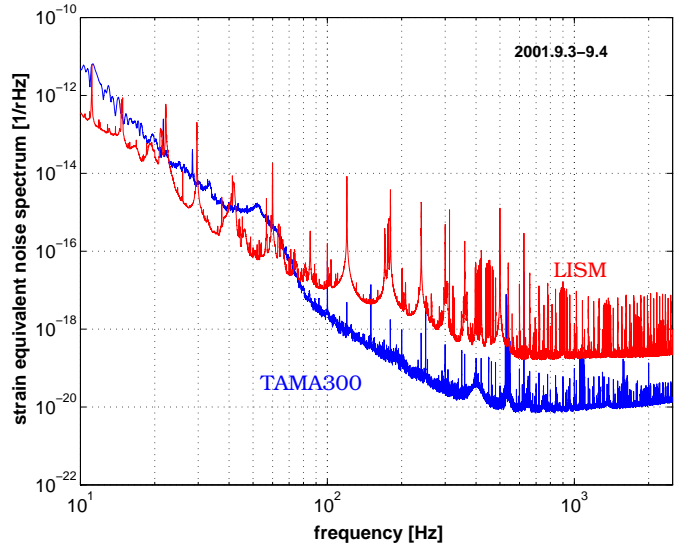}}
\caption{The strain equivalent noise spectra of TAMA300 and LISM 
on September 3, 2001.}
\label{fig:noise}
\end{figure}

In this paper, 
we use the DT6 data taken from September 2nd to 17th, 2001 when
LISM was also in good condition. 
The amount of data available for the coincidence analysis is 275 hours in total. 
Typical one-sided noise power spectra of TAMA300 and LISM 
during this observation are shown in Fig. \ref{fig:noise}. 

\begin{table}[htpd]
\begin{center}
\renewcommand{\arraystretch}{1.5}
\begin{tabular}{c c c}
\hline \hline
    & TAMA300 (DT6)  & LISM   \\
\hline
Interferometer type & Fabry-Perot-Michelson & Locked Fabry-Perot \\
Base length & 300m & 20m \\
Finesse of main cavity & 500 & 25000 \\
Laser Source & Nd:YAG, 10W & Nd:YAG, 700mW \\
Best sensitivity in strain $h\ {\rm[1/\sqrt{Hz}]}$ & $5 \times 10^{-21} 
$ & $6.5 \times 10^{-20}$\\
Location and arm orientation  & $35.68^\circ$N, $139.54^\circ$E,
$225^\circ$ & $36.25^\circ$N, $137.18^\circ$E, $165^\circ$ \\
Maximum delay of signal arrival 
time&\multicolumn{2}{c}{$0.73$msec~~~~~~~~~}  \\
\hline
Operation period& Aug.1 - Sept.20, 2001& Aug.1 - 23, Sept.3 - 17, 2001  
\\
Observation time &1038 hours & 786 hours \\
Operation rate&87\%&91\%\\
\hline
Simultaneous observation &\multicolumn{2}{c}{709 hours~~~~~~~~~}\\
Data used for coincidence analysis &\multicolumn{2}{c}{275 
hours~~~~~~~~~}\\
\hline \hline
\end{tabular}
\end{center}
\caption{Summary of the observation in August and September 2001 by TAMA300 and LISM}
\label{tab:summary}
\end{table}

\begin{table}[htpd]
\begin{center}
\renewcommand{\arraystretch}{1.5}
\begin{tabular}{c c c c c}
\hline \hline
   & Year  & period & obsevation time [hours] &Topics   \\
\hline 
DT1&1999& 6-7 Aug. & 11 &Total detector system check \\
&&&&and  Calibration test\\
DT2&1999& 17-20 Sept.&31& First event search \\
DT3&2000& 20-23 April&13& Sensitivity improved\\
DT4&2000& 21 Aug.-4 Sept.&167 &100 hours observation\\
DT5&2001& 2-10 Mar. &111& Full time observation\\
DT6&2001& 1 Aug.-20 Sept.&1038& 1000 hours observation \\
&&&&and coincident observation with LISM\\
DT7&2002& 31 Aug.-2 Sept.&25& Power recycling installed (Full configuration)\\
DT8&2003& 14 Feb.-14 April & 1158 & Coincident observation with LIGO \\
DT9&2003 - 2004&  28 Nov.- 10 Jan. &  557 & Full automatic operation\\
        &  &  & &and Partial coincident observation \\
        &&&&with LIGO and GEO600 \\
\hline \hline
\end{tabular} 
\end{center} 
\caption{Observation history of TAMA300}  
\label{tab:history} 
\end{table}

\subsection{LISM}\label{sec:detectorl}

LISM is a laser interferometer gravitational wave antenna with arm length of 20m, 
located in the Kamioka mine ($36.25^\circ$N, $137.18^\circ$E), 219.02km west of Tokyo. 
The detector's arm orientation is $165^\circ$ measured counter clockwise from East. 
The LISM antenna was originally developed as a prototype detector from 1991 to 1998 
at the National Astronomical Observatory of Japan, in Mitaka, Tokyo, to demonstrate
advanced technologies \cite{ref:sato}. In 1999, it was moved to the Kamioka mine in order to perform
long-term, stable observations. Details of the LISM detector is found in \cite{ref:satoLISM}. 

The laboratory site is 1000m underground in the Kamioka mine. 
The primary benefit of this location is extremely low seismic noise 
level except artificial seismic excitations. Furthermore, much smaller
environmental variations at this underground site are beneficial to stable operation of 
a high-sensitivity laser interferometer. The optical configuration is the
Locked Fabry-Perot interferometer. The finesse of each arm cavity was
about 25000 to have a cavity pole frequency of 150Hz. 
The main interferometer was illuminated by a Nd:YAG laser yielding 700mW of output power,
and the detector sensitivity spectrum was shot-noise limited at 
frequencies above about 1kHz. 

The operation of LISM was started in early 2000, and has repeatedly been tested and improved since.
The data used in this analysis were taken in the observations between August 1st
and 23th and between September 3rd and 17th, 2001. The total length of data 
is 780 hours. The first half of the period was in a test-run and some
improvements were made after that. The data from the second half were of good
quality to be suitable for a gravitational wave event search, so 323 hours of data for
the latter half was dedicated for this analysis. The best sensitivity during this period
was about $h\sim 6.5 \times 10^{-20}/\sqrt{\rm{Hz}}$ around 800Hz.

\section{Analysis method}\label{sec:method}
\subsection{Matched filtering}
\label{sec:basicformula}

To search for gravitational waves emitted from inspiraling compact binaries,
we use the matched filtering. 
In this method, cross-correlation between observed data and 
 predicted waveforms are calculated to find signals and to estimate
binary's parameters.
When the noise of a detector is Gaussian and stationary, 
the matched filtering is the optimal detection strategy 
in the sense that it gives the maximum
detection probability for a given false alarm probability. 

We use restricted post-Newtonian waveforms as templates: 
the phase evolution
is calculated to 2.5 post-Newtonian order, and the amplitude evolution
is calculated to the Newtonian quadrupole order. 
The effects of spin angular momentum are not taken into account here. 
The filters are constructed in Fourier domain by the stationary phase approximation
\cite{ref:sfa} of the post-Newtonian waveforms \cite{ref:pntemplate}. 
We introduce the normalized templates ${h}_c$ and ${h}_s$ which are 
given in the frequency domain for $f>0$ by 
\begin{eqnarray}
&&\tilde{h}_{c}= N f^{-7/6}\exp(i\Psi(f)), \\
&&\tilde{h}_{s}= iN f^{-7/6}\exp(i\Psi(f)),
\end{eqnarray}
where
\begin{eqnarray}
\Psi(f)&=&2\pi ft_c-{\pi\over 4}+{3\over 128\eta}(\pi G M f c^{-3})^{-5/3}
\Bigg[1+\frac{1}{9}\left(\frac{3715}{84}+55\eta\right)(\pi G M f c^{-3})^{2/3}
-16\pi(\pi M f c^{-3})
\nonumber\\
&&
+\left(\frac{15293365}{508032}+\frac{27145}{504}\eta
+\frac{3085}{72}\eta^2\right)(\pi M f c^{-3})^{4/3}
+\frac{\pi}{3}\left(\frac{38645}{252}+5\eta\right)(\pi M f c^{-3})^{5/3}\Bigg],
\label{eq:2.5PN}
\end{eqnarray}
where $f$ is the frequency of gravitational waves, $t_c$ is the coalescence time,
$M=m_1+m_2$, $\eta=m_1m_2/M^2$, and $m_1$ and $m_2$ are the masses of
binary stars.
For $f<0$, they are given by $\tilde{h}_{c/s}(f)=\tilde{h}_{c/s}^*(-f)$,
where the asterisk denotes the complex conjugation. 
The normalization factor $N$ is defined such that $h_{c}$
and $h_{s}$ satisfy
\begin{eqnarray}
&& ({h}_c,{h}_c)=1, \quad\quad ({h}_s,{h}_s)=1,
\end{eqnarray}
where
\begin{eqnarray}
(a,b)\equiv 2\int^{\infty}_{-\infty}df
{\tilde{a}(f)\tilde{b}^*(f)\over S_n(|f|)}.
\label{eq:innerproduct}
\end{eqnarray}
$S_n(f)$ is the strain equivalent 
one-sided noise power spectrum density of a detector.
We note that, for $\tilde{h}_c$ and $\tilde{h}_s$ 
calculated by the stationary phase approximation, 
we have $({h}_c,{h}_s)=0$. 

In the matched filtering, we define the filtered output by 
\begin{eqnarray}
\tilde{\rho}=(s,{h}_c \cos(\phi_c)+{h}_s \sin(\phi_c)),
\end{eqnarray}
where $s(t)$ is the signal from a detector and 
$\phi_c$ is the phase of the template waveform. 
For a given interval of $t_c$, we maximize $\tilde{\rho}$ 
over the parameters $t_c, $$M$, $\eta$ and $\phi_c$. 
The filtered output maximized over $\phi_c$ is given by 
\begin{eqnarray}
\tilde{\rho}&=&\sqrt{(s,{h}_c)^2+(s,{h}_s)^2}\equiv \rho.
 \label{eq:rho} 
\end{eqnarray}
The square of the filtered output, $\rho^2$, has an expectation value $2$
in the presence of only Gaussian noise in the data $s(t)$. 
Thus, we define the signal-to-noise ratio, SNR,
by $\rho/\sqrt{2}$. 

Matched filtering is the optimal detection strategy 
in the case of stationary and Gaussian noise of detector.
However, since the detectors' noise is not stationary and Gaussian in the case of real 
laser interferomters, we introduce $\chi^2$ method to the matched filtering 
in order to discriminate such noise from real gravitational wave signals. 
We describe details of $\chi^2$ method in Appendix~\ref{sec:Ap1}. 

\subsection{Algorithm of the matched filtering analysis} 
\label{sec:matchedfiltering}

In this subsection, we describe a method to analyze
time sequential data from the detectors by matched filtering. 

First, we introduce, ``a continuously locked segment''.
The TAMA300 and LISM observations were sometimes interrupted by 
the failure of the detectors to function normally, which are usually 
called ``unlock'' of the detectors, or were interrupted 
manually in order to make adjustments to the instruments. 
A continuously locked segment is a period
in which the detector is continuously operated
without any interruptions and the data 
is taken with no dead time. 
In the analysis of this paper, we treat only the data in
such locked segments.

The time sequential voltage data of a continuously locked segment
are divided into small subsets of data with length of
52.4288 seconds (= sampling interval [s] $\times$ number of samples
= $(5\times 10^{-5})\times 2^{20}$ [s]).
Each subset of data has overlapping portions with adjacent 
subsets 
for 4.0 seconds in order not to lose signals which lie across
borders of two adjacent subsets. 
The data of a subset are Fourier transformed into frequency domain
and are multiplied by 
the transfer function to transform into strain equivalent data. 
The resulting subset of data is the signal of the detector 
in the frequency domain, $\tilde{s}(f)$, used in the matched filtering. 

The power spectrum density of noise $S_n(f)$ is basically
evaluated in a subset of data neighbouring to each $s(t)$ 
except for the cases below. 
On estimating the noise power spectrum, $S_n(f)$, we do not use the data 
contaminated by transient burst noise. For this purpose,
we evaluate the fluctuation of 
the noise power defined by 
\begin{equation}
p=\left[4\int_0^\infty{f^{-7/3}\over S_n(f)}df\right]^{-1/2},
\end{equation}
for each set of data with length of 65.6 seconds
which composes one file of stored data.
We also calculate the average of $p$, 
$\langle p\rangle$, within each continuously locked segment. 
For each $\tilde{s}(f)$, we then apply the following criterion. 
If a subset of data in the neighborhood
of $\tilde{s}(f)$ lies entirely in one of the files, we examine the value of $p$ of
the file, and if it deviates from the average $\langle p\rangle$ for more 
than 2dB, i.e., $p > 1.26 \langle p\rangle$, 
we do not use that subset of data for evaluating the power spectrum
and move to the neighboring subset.
If a neighboring subset lies over two files, we examine the values of $p$ of
the two files, and if either of them exceeds the 2dB level, we 
use neither of them. 
If a neiboring subset such that a file (or two consecutive files)
that contains it has $p < 1.26 \langle p\rangle$ is found, 
the subset is divided into 8 pieces and
the $S_n(f)$ is evaluated by taking the average of them. 
If the fluctuations of $p$ are too large, and we cannot find
files with the values of $p$ within 2dB of the average
within the locked segment,
we use the power spectrum which is evaluated by taking the average of 
all the data in the corresponding locked segment.


In order to take the maximization of $\rho$ in Eq.(\ref{eq:rho}) 
over the mass parameters, 
we introduce a grid in the mass parameter space. 
Each grid point defines the mass parameters which characterize
a template. 
We adopt the algorithm introduced in \cite{ref:TT}
to define the grid point in the mass parameter space. 
The distance between the grid points 
is determined 
so as not to lose more than 3 $\%$ of signal-to-noise ratio
due to mismatch between actual mass parameters 
and those at grid points. 
Accordingly, the mass parameter space depends on the power spectrum
of noise. In order to take into account of the changes in
the noise power spectrum with time,
we use different mass parameter spaces for different locked segments.
For each locked segment, the averaged power spectrum of noise 
is used to determine the grid spacing in the mass parameter space.

We consider the mass of each component star in the range
$1M_\odot\leq m_1,\,m_2\leq 2M_\odot$. 
This mass range is chosen so that it covers the
most probable mass of a neutron star, $\sim 1.4 M_\odot$. 

With $\tilde{s}(f)$, $S_n(f)$ and a template on each grid point
of the mass parameter space,
we calculate $\rho$ in Eq. (\ref{eq:rho}). 
For each interval $\Delta t= 25.6$ msec, 
we search for $t_c$ at which the local maximum of $\rho$ is realized. 
If the $\rho$ thus obtained is greater than a pre-determined value
$\rho_m$, we calculate the value of $\chi^2$
as discussed in Appendix~\ref{sec:Ap1}. 
We adopt $\rho_m=7$ in this paper. Choosing a too large
$\rho_m$ results in missing actual events from the data, while
a too small $\rho_m$ requires too much computational time.
The same computation is done for all the mass parameters on each grid point. 

Finally, for each interval of the coalescence time
with length $\Delta t= 25.6$ msec, 
we search for $t_c$, $M$, $\eta$ 
which realize the local maximum of $\rho$. 
Each maximum is considered a event. 
The value of $t_c$, $\rho$, $\chi^2$, $M$, $\eta$ 
of each event are recorded in event lists. 

\section{Results of matched filtering search}
\label{sec:matchedfilterresults}

In this section, we show 
the result of the independent analysis for each detector. 

Our analysis is carried out with 9 Alpha computers and also with 12 
Pentium4 computers at Osaka University. 
The matched filtering codes are paralleled
by the MPI library. 
Among the data from September 3rd to 17th, 2001,
TAMA300 has 292.4 hours of data after removing unlocked periods.
We also removed the data segments of lengths less than 10 minutes. 
The total length of data is 287.6 hours. 
LISM has 323.0 hours of data after removing unlocked periods.
After removing the data segments less than 10 minutes, 
the total length of data is 322.6 hours. 

\begin{figure}
\scalebox{0.5}[0.5]{\includegraphics{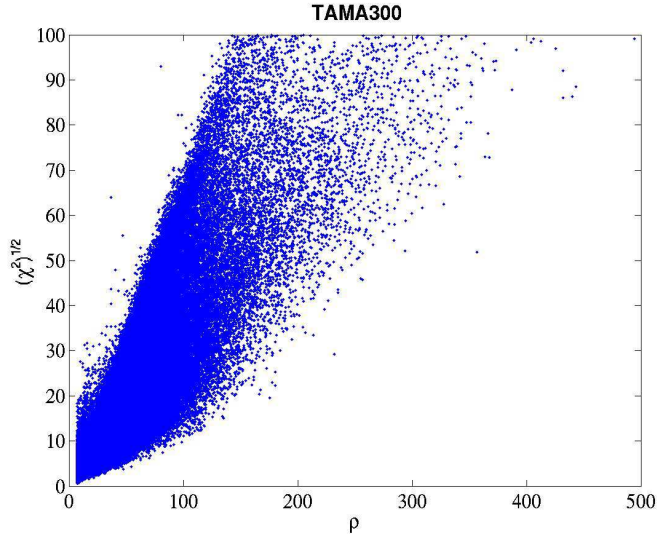}}
\caption{Scatter plots ($\rho , \sqrt{\chi^2}$)
of the events of TAMA300.}
\label{fig:tamacom25}
\end{figure}
\begin{figure}
\scalebox{0.5}[0.5]{\includegraphics{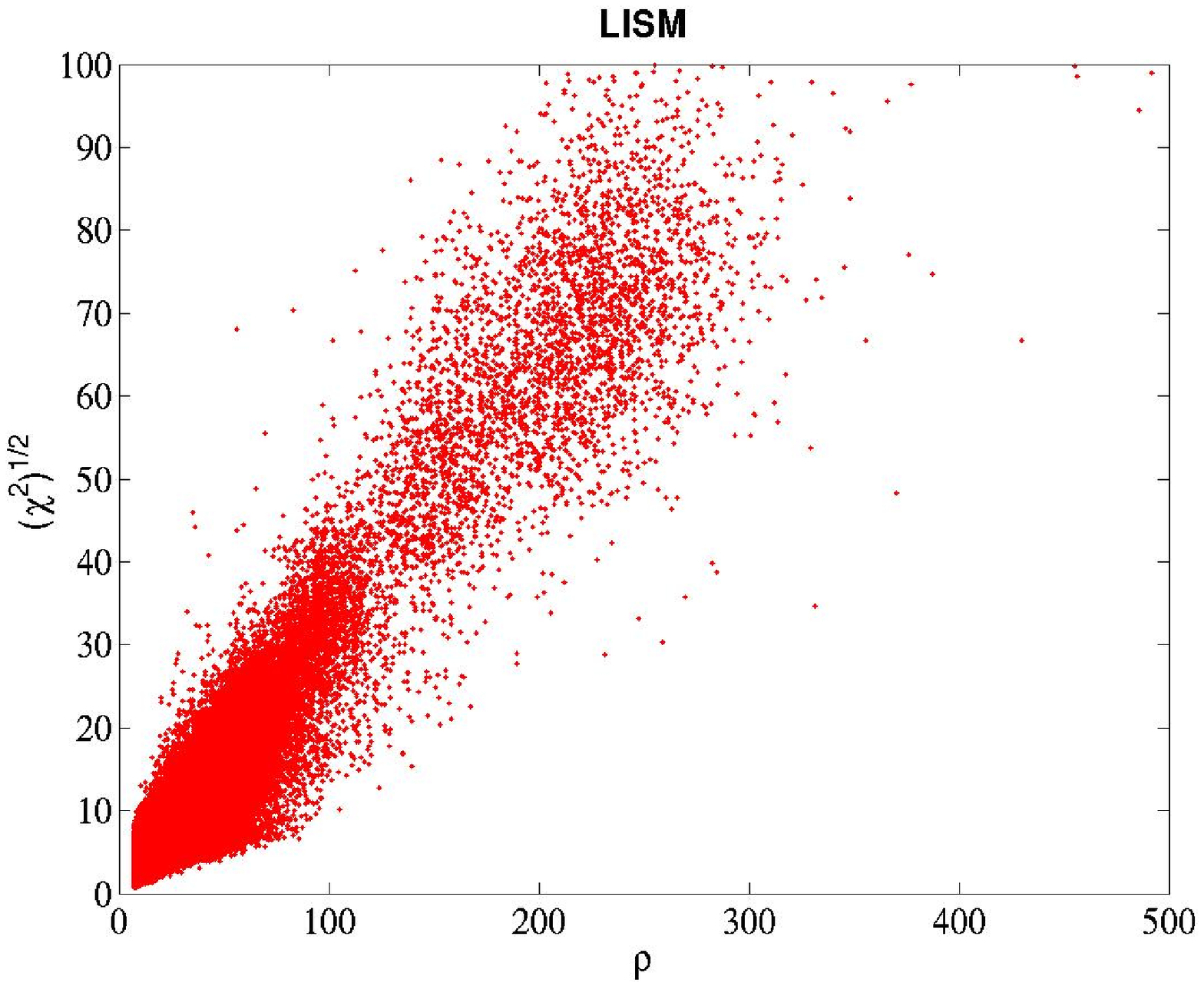}}
\caption{The same figure as Fig.\ref{fig:tamacom25} but for LISM.}
\label{fig:lismcom25}
\end{figure}
\begin{figure}
\scalebox{0.5}[0.5]{\includegraphics{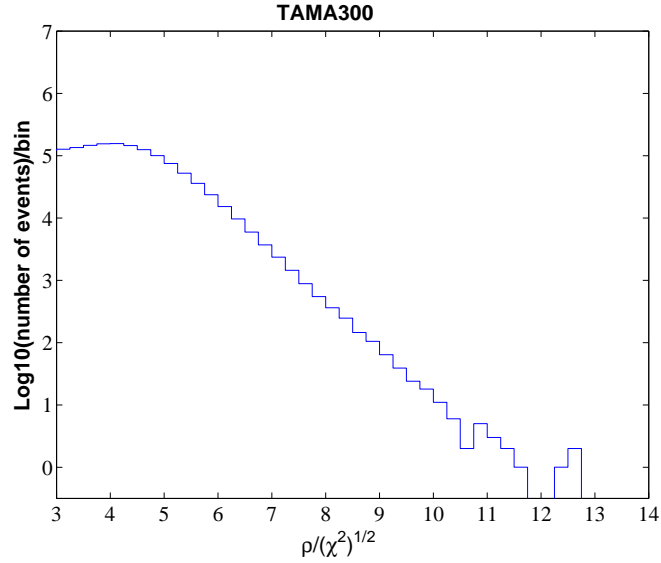}}
\caption{Histogram of the number of events 
of TAMA300 in terms of $\rho / \sqrt{\chi^2}$. }
\label{fig:histcomTAMA25}
\end{figure}
\begin{figure}
\scalebox{0.5}[0.5]{\includegraphics{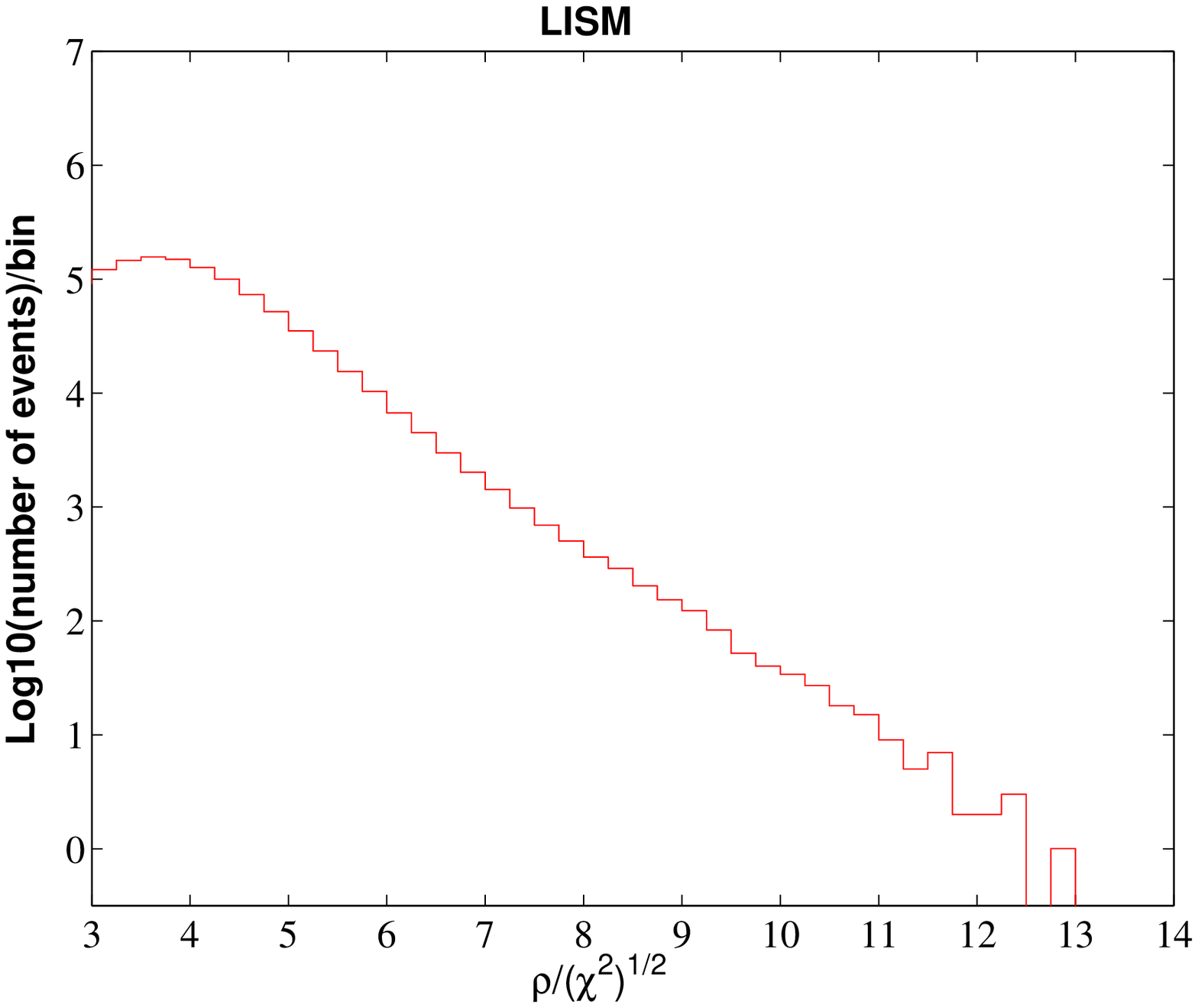}}
\caption{The same figure as Fig.\ref{fig:histcomTAMA25} but for LISM.}
\label{fig:histcomLISM25}
\end{figure}


The scatter plots of $(\rho , \sqrt{\chi^2})$ of the
events are shown in Figs. \ref{fig:tamacom25} and \ref{fig:lismcom25}.
We discriminate the non-Gaussian noise from real gravitational
wave signals by setting the threshold to the value of $\rho / \sqrt{\chi^2}$ 
(see Appendix~\ref{sec:Ap1}). 
In Figs. \ref{fig:histcomTAMA25} and \ref{fig:histcomLISM25} , 
we show the number of events for bins of $\rho / \sqrt{\chi^2}$. 

Although the main topic of this paper is to perform 
a coincidence analysis, 
for the purpose of comparison between 
a single-detector analysis and a coincidence analysis, 
we evaluate the upper limit to the 
event rate which is derived from an analysis independently done
for each detector.
The upper limit to the Galactic event rate is calculated by \cite{ref:40m} 
\begin{equation}
R = \frac{N}{T \epsilon} \label{eq:eventrate}
\end{equation}
where $N$ is the upper limit to the average number of
events with $\rho/\sqrt{\chi^2}$ greater than a pre-determined 
threshold, $T$ is the total length of data [hours] and 
$\epsilon$ is the detection probability.

\begin{figure}
\scalebox{0.5}[0.5]{\includegraphics{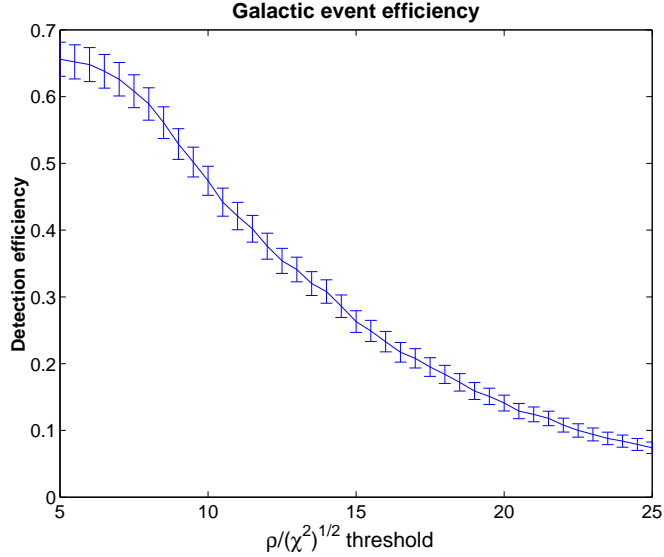}}
\caption{Galactic event detection efficiency of TAMA300. 
The error bars shows the $1\sigma$ error of the simulation.}
\label{fig:efficiencyTAMA}
\end{figure}

To examine the detection probability of the Galactic neutron 
star binary events, 
we use a model of the distribution of neutron star binaries in our Galaxy
which is given by \cite{ref:neutronstar}
\begin{equation}
dN = e^{-R^2 / 2R _0 ^2} e^{-Z / h_z} R dR dZ ,
\end{equation}
where $R$ is Galactic radius, $R_0 = 4.8$ kpc, 
$Z$ is height off the Galactic plane and $h_z = 1$ kpc is the scale height.  
We assume that the mass distribution is uniform
between $1M_{\odot}$ and $2 M_{\odot}$. 
We also assume uniform distributions for the inclination angle
and the phase of an event.
With these distribution functions, we 
perform a Monte Carlo simulation.
The simulated gravitational wave events
are injected into the data of each detector for about every 15 minutes. 
We perform a search using the same code used in 
our matched filter analysis, and evaluate the detection probability
for each $\rho/\sqrt{\chi^2}$ threshold. 
The result for TAMA300 
is shown in Fig.~\ref{fig:efficiencyTAMA}. 

\begin{figure}
\scalebox{0.5}[0.5]{\includegraphics{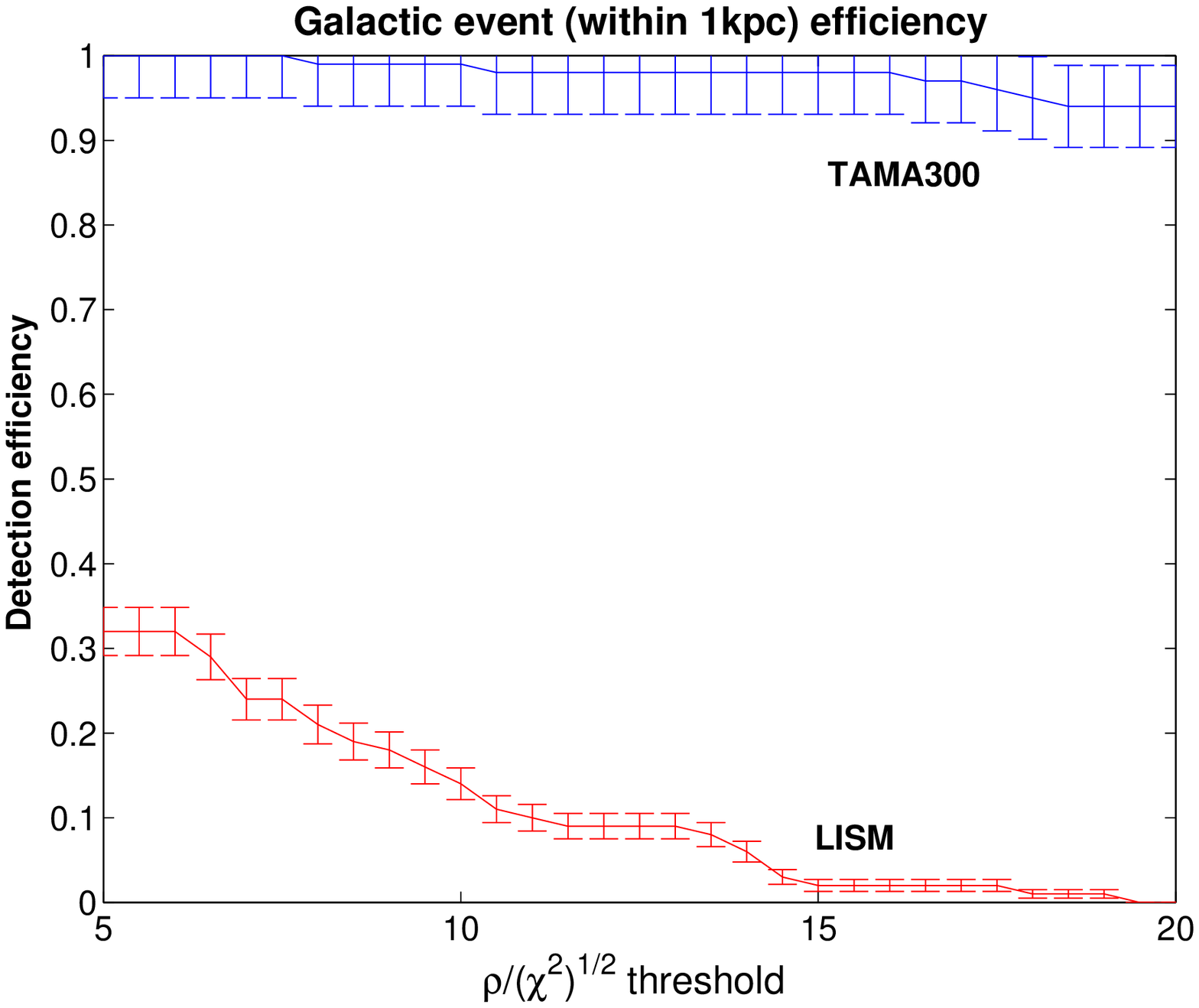}}
\caption{The detection efficiency of TAMA300 and LISM for
nearby events within 1kpc.
The error bars show the $1\sigma$ error of the simulation.}
\label{fig:efficchiTAMALISM}
\end{figure}

For the case of LISM, since LISM's sensitivity is
not good enough to observe events in all of the Galaxy,
we only evaluate 
the detection probability of nearby events within 1kpc. 
The result is shown in Fig.~\ref{fig:efficchiTAMALISM}.

The threshold of $\rho/\sqrt{\chi^2}$ for each of the analysis
is determined by the fake event rate. 
We set the fake event rate to be $2.0$\,[1/yr]. 
We approximate the distribution of $\rho/\sqrt{\chi^2}$ 
in each of Figs.~\ref{fig:histcomTAMA25} and \ref{fig:histcomLISM25}
by an exponential function and extrapolate it to
large $\rho/\sqrt{\chi^2}$.
We assume that 
this function describes the background fake event distribution. 

For the TAMA300 case, 
the fake event rate $N_{\rm bg}/T=2.0$ [1/yr]\,$=0.00023$\,[1/hour]
gives the total number of expected fake events as $N_{\rm bg}=0.066$. 
This determines the threshold to be $\rho / \sqrt{\chi^2} = 14.8 $. 
With this threshold, we obtain the detection probability, $\epsilon=0.263$,
from Fig.~\ref{fig:efficiencyTAMA}. On the other hand,
the number of observed events with $\rho/\sqrt{\chi^2}$ greater than the threshold is $N_{\rm obs}=0$. 
Using Bayesian statistics, and assuming 
uniform prior probability for the real event rate and 
the Poisson distributions for real and background events, 
we estimate the expected number of real events $N$ which $\rho/\sqrt{\chi^2}$ is 
greater than the threshold 
with a given confidence level (CL).
Namely, it can be evaluated from 
the equation \cite{ref:basian}, 
\begin{equation}
{{e^{-(N+N_{\rm bg})}\sum_{n=0}^{n=N_{\rm obs}}{(N+N_{\rm bg})^n\over n!}}\over 
{e^{-N_{\rm bg}}\sum_{n=0}^{n=N_{\rm obs}}{(N_{\rm bg})^n\over n!}}}
=1-\rm{CL} \,. \label{eq:upperlimitformula}
\end{equation}
Using this formula, we obtain the upper limit to 
the expected number of real events to be 2.30 with $90\%\,\rm{CL}$.
Then, using the length of data $T=287.6$ hours, we obtain 
the upper limit of the event 
rate as $R_{90\%} = 0.030$\,[1/hour] ($\rm{CL}=90\%$). 

For the LISM detector, we only evaluate
the upper limit to nearby events within 1kpc. 
We set the threshold $\rho / \sqrt{\chi^2} = 14.6$, 
corresponding to the number of expected fake events $N_{\rm bg}=0.074$ 
which realizes the fake event rate 
$N_{\rm bg}/T=2.0$\,[1/yr]. 
The number of observed events with $\rho/\sqrt{\chi^2}$ greater than the threshold is $N_{\rm obs}=0$. 
Thus, the upper limit to the expected number of real events is 
again $2.30$ with $90\%$ CL. 
The detection probability is given from Fig.\ref{fig:efficchiTAMALISM}  
as $\epsilon=0.042$. The length of data is $T=322.6$\,hours. 
Using these numbers, we obtain the upper limit to the nearby event rate
as $0.17$\,[1/hour] with $90\%\,\rm{CL}$.

The results of matched filtering analysis for TAMA300 and LISM
are summarized in Table \ref{tab:matched-results}. 

\begin{table}[htpd]
\begin{center}
\renewcommand{\arraystretch}{1.5}
\begin{tabular}{c c c c c c}
\hline \hline
   &threshold&N&Detection efficiency&Length of data &Upper limit (90\% CL)\\
\hline 
TAMA300 &14.8&2.30 (90\% CL)  &0.263&287.6 [hours]&0.030 [1/hour]\\
LISM    &14.6&2.30 (90\% CL)  &0.042&322.6 [hours]&0.17  [1/hour] (for nearby events) \\
\hline \hline
\end{tabular} 
\end{center} 
\caption{Results of matched filtering analysis for TAMA300 and LISM}  
\label{tab:matched-results} 
\end{table}

\section{Coincidence analysis}
\label{sec:coincidence}

\subsection{Method}

In the previous section, we obtained 
event lists for TAMA300 and LISM. 
Each event is characterized by $t_c$, ${\cal M}$, $\eta$,
$\rho$, and $\chi^2$, where ${\cal M}$ is the chirp mass
($=M \eta^{3/5}$).
True gravitational wave events will 
appear in both event lists with different values of these
parameters according to the detectors' noise, the difference
in the detectors' locations and thier arm orientations, 
and the discreteness of the template space.
In this section, we evaluate the difference of the parameters real events have. 


\vspace{0.5cm}
{\it Time selection}: 

The distance between the TAMA300 site and the LISM site is 219.02km. 
Therefore, the maximum delay of the arrival time of 
gravitational wave signals is $\Delta t_{\rm dist} =0.73057$\,msec.
The allowed difference in $t_c$ is set as follows. 
If the parameter,
 $t_{c,{\mbox{\rm\tiny TAMA}}}$ and $t_{c,{\mbox{\rm\tiny LISM}}}$, 
of an event satisfy
\begin{equation}
|t_{c,{\mbox{\rm\tiny TAMA}}}-t_{c,{\mbox{\rm\tiny LISM}}}| < 
\Delta t_{\rm dist}+\Delta t_{\rm noise},
\end{equation}
the event is recorded in the list as a candidate for real events.
We estimate errors in $t_c$ due to noise $\Delta t_{\rm noise}$
by using the Fisher information 
matrix (see Appendix \ref{sec:Ap2} for 
a detailed discussion). 
We denote the 1 $\sigma$ value of the error of $t_c$ by
$\Delta t_{c,i}$ for $i=$ TAMA or LISM.  
We determine $\Delta t_{\rm noise}$ as 
$\Delta t_{{\rm noise}}=\sigma_w \times \Delta t_c$
where $\Delta t_{c}=\sqrt{\Delta t^2_{c,{\mbox{\rm\tiny TAMA}}}+
\Delta t^2_{c,{\mbox{\rm\tiny LISM}}}}$. The parameter $\sigma_w$ is 
to be determined in such a way that it is small enough to
exclude accidental coincidence events effectively
but is large enough to make the probability for missing a
real event sufficiently small.

In this paper, we adopt $\sigma_w=3.29$ which corresponds
to 0.1$\%$ probability of losing real signals 
if the noise are Gaussian and if both detectors 
are located at the same site. 
Although it may be possible to tune the value of $\sigma_w$ to obtain 
a better detection efficiency 
while keeping the fake event rate low enough,
we do not bother to do so.
Instead, we check whether we have a reasonable detection efficiency 
by this choice. 
To check the detection efficiency is important in any case,
since the $\Delta t_c$ determined above 
assumes a large signal amplitude in the presence of
Gaussian noise. The actual detection efficiency might
 be different from what we expected. 

\vspace{0.5cm}
{\it Mass selection}:

In the same way as for $t_c$,  
errors in the values of $\mathcal{M}$ and $\eta$ 
due to detector noise, $\Delta \mathcal{M}_{\rm noise}$ and
$\Delta \eta_{\rm noise}$, are estimated by using
the Fisher matrix. 
We denote the 1 $\sigma$ values of errors in
$\mathcal{M}$ and $\eta$ by
$\Delta \mathcal{M}_i$ and $\Delta \mathcal{\eta}_i$,
respectively.
We set
$\Delta \mathcal{M}_{\rm noise}
=\sigma_w \sqrt{(\Delta \mathcal{M})^2_{\mbox{\rm\tiny TAMA}}+
(\Delta \mathcal{M})^2_{\mbox{\rm\tiny LISM}}}$ and 
$\Delta \eta_{\rm noise}=\sigma_w \sqrt{(\Delta \eta)^2_{\mbox{\rm\tiny TAMA}}+
(\Delta \eta)^2_{\mbox{\rm\tiny LISM}}}$,
and adopt $\sigma_w=3.29$ as in the case of $t_c$. 

When the amplitude of a signal is very large, 
errors due to detector noise become small since
they are inversely proportional to $\rho$,
and errors due to the discreteness of the mass
parameter space become dominant.
We denote the latter errors by
$\Delta \mathcal{M}_{\rm mesh}$ and $\Delta \eta_{\rm mesh}$.
They are determined from the maximum 
difference in the neighbouring 
mesh points in the mass parameter space. 

By taking account of the above two effects, 
we choose the allowable difference in the mass parameters as 
\begin{eqnarray}
|\mathcal{M}_{\rm TAMA} - \mathcal{M}_{\rm LISM}| &<& 
\Delta \mathcal{M}_{\rm noise}+\Delta \mathcal{M}_{\rm mesh} , \\
|\eta_{\rm TAMA} - \eta_{\rm LISM}| &<& \Delta \eta_{\rm noise} 
+\Delta \eta_{\rm mesh}. 
\end{eqnarray}

\vspace{0.5cm}
{\it Amplitude selection}:

Since the two detectors have different sensitivities,
signal-to-noise ratios of an observed gravitational wave
signal will be different for the two detectors. 
Further, since their arm orientations are different, 
the signal-to-noise ratios will differ
even if they have the same noise power spectrum. 

We express the allowable difference in $\rho_{\mbox{\rm\tiny TAMA}}$ and 
$\rho_{\mbox{\rm\tiny LISM}}$ as
\begin{equation}
\delta _{\rm sens} - \delta _{\rm arm} -\delta_{\rm noise}\le 
\log
\Big({\frac{\rho_{\mbox{\rm\tiny TAMA}}}{\rho_{\mbox{\rm\tiny LISM}}}}\Big) 
\le \delta _{\rm sens} + \delta _{\rm arm}+\delta_{\rm noise}. 
\end{equation}
Here, $\delta_{\rm sens}$ is due to the difference in $S_n$,
\begin{equation}
\delta _{\rm sens}\equiv 
\log
\Big[\Big(\int
 \frac{f^{-7/3}}{S_{n\ {\mbox{\rm\tiny TAMA}}}(f)} df \Big)^{1/2}/
\Big(\int \frac{f^{-7/3}}{S_{n\ \mbox{\rm\tiny LISM}}(f)} df \Big)^{1/2}
  \Big],
\end{equation}
and $\delta_{\rm arm}$ is due to the difference in the arm orientations, 
and $\delta_{\rm noise }$ is due to detector noise. 
The value of $\delta_{\rm noise}$ is evaluated by the Fisher matrix in the same way as
$t_c$ and masses. 

The value of $\delta_{\rm sens}$ is determined for each event
individually from the noise power spectrum used in the matched filtering.
$\delta_{\rm arm}$ is evaluated by a Monte Carlo simulation
as follows.
We assume that the two detectors have the same noise power spectrum,
and generate the waveforms of Galactic events randomly.
We then evaluate $\rho$ of all the events detected by each detector,
and determine the value of $\delta_{\rm arm}$ in such a way
that for more than 99.9 \% of events, we have 
$|\log(\rho_{\mbox{\tiny TAMA}}/\rho_{\mbox{\tiny LISM}})|
\leq \delta_{\rm arm}$. 
This gives $\delta_{\rm arm}=1.60$. 

\subsection{Detection efficiency and the parameter windows}
\label{sec:efficiency}

Here, we discuss the detection efficiency of our coincidence analysis. 
In particular, we examine the validity of the choice $\sigma_w=3.29$ 
made in the previous section.

\begin{figure}
\scalebox{0.5}[0.5]{\includegraphics{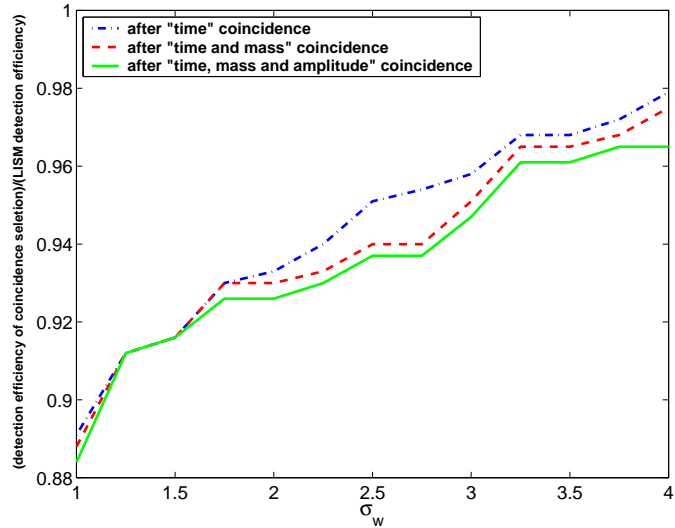}}
\caption{Relative detection efficiency of the coincidence analysis compared 
to the single-detector efficiency of LISM as a function of 
the parameter $\sigma_w$ used for the coincidence criterion. 
The dot-dashed line is the efficiency after the time
selection, the dashed line is the efficiency after the time and mass
selection, and the solid line is the efficiency after the 
time-mass-amplitude
selection.}
\label{fig:coineffic}
\end{figure}

For the Galactic event simulation
discussed in Section~\ref{sec:matchedfilterresults},
the detection 
efficiencies of TAMA300 and LISM for the threshold $\rho/\sqrt{\chi^2}>7$
are 99\% and 24\%, respectively. 
The detection efficiency of the coincidence analysis 
is dominated by the LISM's efficiency.  
Thus we define the detection efficiency for the coincidence analysis,  
as the fraction of LISM events which fulfill the coincidence criteria.
The result is shown in Fig.~\ref{fig:coineffic}.
We find that 
 more than 94 \% of LISM events can be detected if we set $\sigma_w>3$.
Thus with $\sigma_w=3.29$, we have a reasonably high detection
efficiency.

If we adopt a larger value of $\sigma_w$, 
we obtain a higher detection efficiency,
but the number of fake events will also increase,
and vise versa for a smaller value of $\sigma_w$.
Then, one may tune the value of $\sigma_w$ so that
it gives the most stringent upper limit to the event rate. 
However, since we cannot expect any drastic improvement
by such an optimization, we adopt $\sigma_w=3.29$
in this paper for the sake of simplicity of the analysis.

\subsection{Results}
\label{sec:coincidenceresults}

In this subsection we discuss the results of the coincidence analysis.
The length of data used for the coincidence analysis is 275.3 hours when
both TAMA300 and LISM detector were operated simultaneously. 

\begin{figure}
\scalebox{0.5}[0.5]{\includegraphics{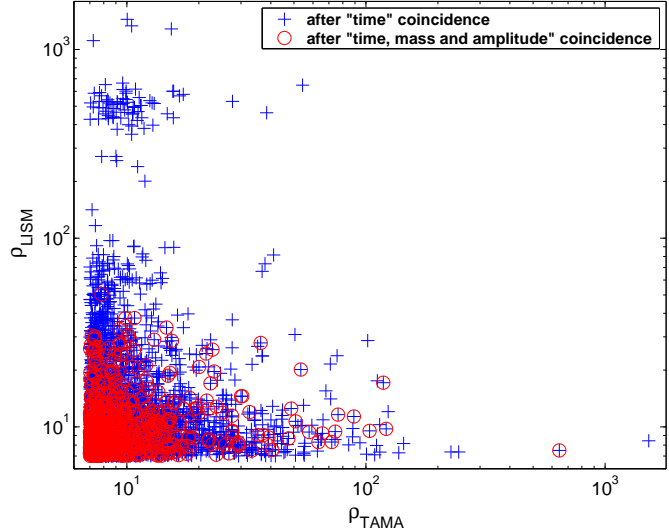}}
\caption{($\rho_{\mbox{\rm\tiny TAMA}}$ ,$\rho_{\mbox{\rm\tiny LISM}}$)
scatter plots.The crosses (+)
are the events survived after the time selection, 
and the circled crosses ($\protect\oplus$)
 are the events survived after the time, mass and amplitude 
selections.}
\label{fig:scatterrhocoin25}
\end{figure}

As a result of independent matched filtering searches, 
we obtained 1,868,388 events from the TAMA300 data 
and 1,292,630 events from the LISM data. 
For these events, we perform the time, mass 
and amplitude selections discussed in the previous section. 
In Fig.~\ref{fig:scatterrhocoin25}, 
we show a scatter plot of the events after coincidence 
selections in terms of 
$\rho_{\mbox{\rm\tiny TAMA}}$ and $\rho_{\mbox{\rm\tiny LISM}}$. 
A significant number of events are removed by imposing 
coincidence conditions. 
Only $0.04\%$ of the TAMA300 events remain. 
In Table \ref{tab:coincident-results}, we show the
number of events which survived after the selections.

\begin{figure}
\scalebox{0.5}[0.5]{\includegraphics{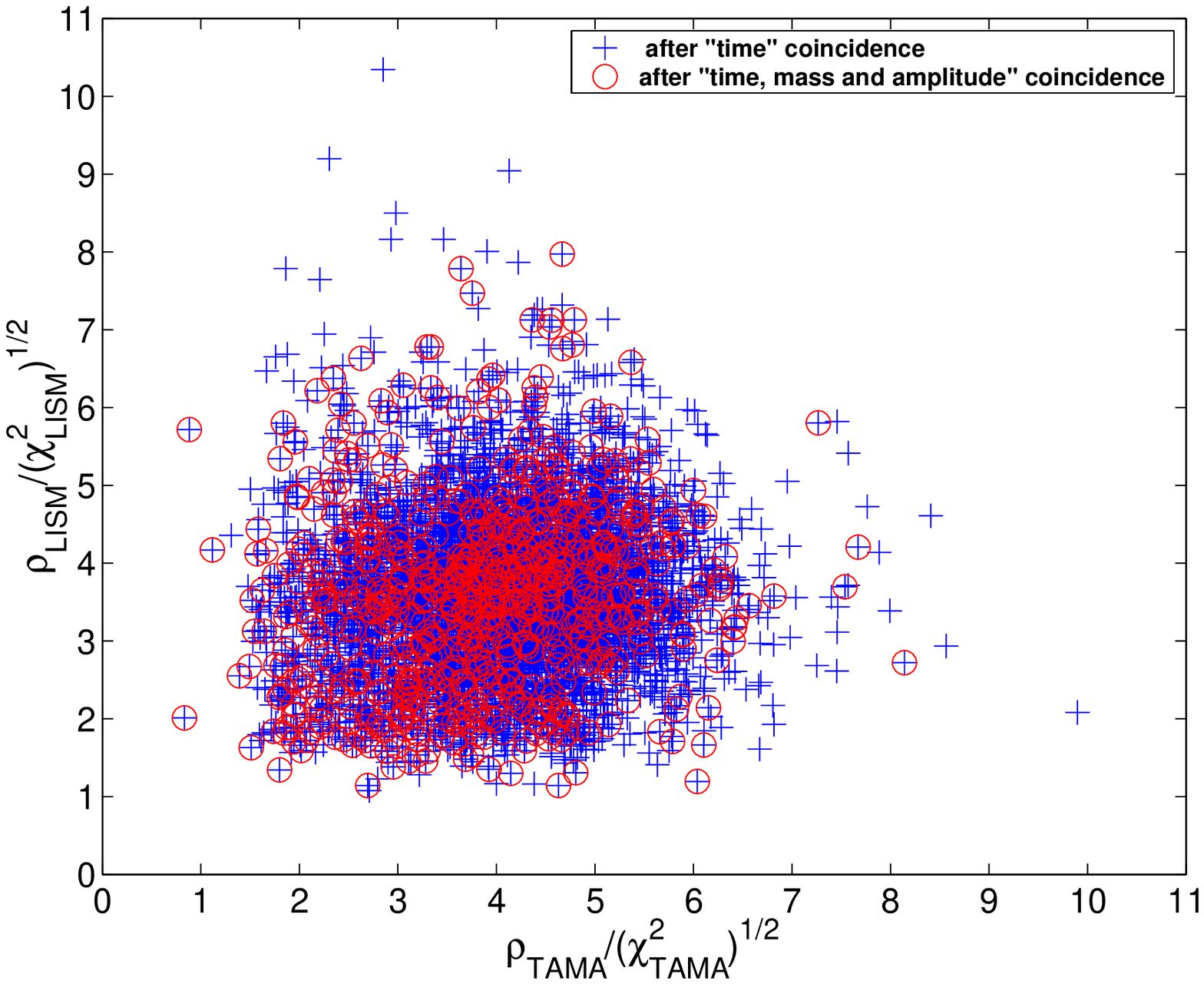}}
\caption{($\rho_{\mbox{\rm\tiny TAMA}}/\sqrt{\chi^2_{\mbox{\rm\tiny TAMA}}}$ 
,$\rho_{\mbox{\rm\tiny LISM}}/\sqrt{\chi^2_{\mbox{\rm\tiny LISM}}}$)
scatter plots.The crosses (+)
are the events survived after the time selection, 
and the circled crosses ($\protect\oplus$)
 are the events survived after the time, mass and amplitude 
selections.}
\label{fig:determthre25}
\end{figure}

We reduce the fake events by introducing the renormalization $\rho$ by $\chi^2$
in addition to the coincidence conditions.
In Fig.~\ref{fig:determthre25}, 
we show a scatter plot of these 
events in terms of the value of 
$\rho_{\mbox{\rm\tiny TAMA}}/\sqrt{\chi^2_{\mbox{\rm\tiny TAMA}}}$ 
and $\rho_{\mbox{\rm\tiny LISM}}/\sqrt{\chi^2_{\mbox{\rm\tiny LISM}}}$.


\begin{table}
\begin{center}
\renewcommand{\arraystretch}{1.5}
\begin{tabular}{c c c}
\multicolumn{3}{c}{Results of independent matched filtering searches}
\\
\hline \hline
 &TAMA300&LISM\\
\hline
Number of events & 1,868,388 & 1,292,630\\
\hline \hline
\\
\multicolumn{3}{c}{Results of coincidence analysis}
\\
\hline \hline
   & $n_{\rm obs}$ & $\bar{n}_{\rm acc} \pm \bar{\sigma}_{\rm acc}$    \\
\hline
after time selection & 4706   & $(4.2 \pm 0.5)\times 10^3$ \\
after time and mass selection & 804  & $(7.1 \pm 0.8)\times 10^2$   \\
after time, mass and amplitude selection & 761 & $ (6.7 \pm 0.8)\times 10^2 $
  \\
\hline \hline
Threshold & $N_{\rm obs}$ &$N_{\rm bg}$\\
\hline
$\rho_{\mbox{\rm\tiny TAMA}}/\sqrt{\chi^2_{\mbox{\rm\tiny TAMA}}}>8.3$ 
and
$\rho_{\mbox{\rm\tiny LISM}}/\sqrt{\chi^2_{\mbox{\rm\tiny LISM}}}>8.1$
&0 &0.063\\
\hline \hline
\end{tabular} 
\end{center} 
\caption{Results of coincidence analysis. 
$n_{\rm obs}$ is the number of coincidence events. 
$\bar{n}_{\rm acc}$ and $\bar{\sigma}_{\rm acc}$ are the 
estimated number of accidental coincidence and its variance,
respectively.
Note that the mean number of accidentals and their variance
after the time selection procedure affect those
after the time and mass selection procedure, and
the latter affect those after the time, mass and amplitude
selection procedure.
Thus, because the observed number of coincidence events is 
consistent with the expected number of accidental coincidence
after the time selection procedure, it is not unnatural
to find a good agreement
between the observed value and the expectation value
in each of the subsequent selection procedures.}  
\label{tab:coincident-results} 
\end{table}


In order to obtain statistical significance from the above results, 
the number of coincident events should be compared with the number
 of accidental 
coincidences produced purely by noise events. 
If events occur completely randomly, and its event 
rate in each detector is stationary, the average 
number of accidental coincidences after the time selection is given by 
\begin{equation}
\bar{n}_{\rm pr} = N_{\mbox{\rm\tiny TAMA}} N_{\mbox{\rm\tiny LISM}} 
\frac{\Delta \bar{t_{c}}^{\rm window}}{T_{\rm obs}} , \label{eq:pr}
\end{equation}
where $N_{\mbox{\rm\tiny TAMA}}$ and $N_{\mbox{\rm\tiny LISM}}$ 
are the number of events in each detector, 
$T_{\rm obs}$ is the total observation time,
 and $\Delta \bar{t}_{c}^{\rm window}$ is 
the averaged value of the time selection window. 
The averaged value of the time selection window is 
evaluated as $\Delta \bar{t}_{c}^{\rm window}=1.29$ msec. 
We thus obtain $\bar{n}_{\rm pr} = 6.3\times 10^3$, 
which is slightly larger than the observed number of coincidence, 4706,
 after the time selection.
One reason for this diffrence is that the event
 trigger rate is not stationary over the whole
period of this observation. 

In order to obtain a more reliable value for the rate of
accidental coincidence, 
we use the time shift procedure. 
Namely, we shift all events of one detector by a time $\delta t$ artificially 
(which is called the time delay), and perform coincidence searches
to determine the number of accidental events
$n_{c}(\delta t)$ for various values of $\delta t$ \cite{ref:Amaldi} \cite{ref:Astone}.
With $m$ different values of time delay, 
we calculate the expected number of coincident events and its standard 
deviation as 
\begin{equation}
\bar{n}_{\rm acc} = \frac{1}{m}\sum_{i=1}^{m} n_{c}(\delta t_{i}), \label{eq:acc}
\end{equation}
\begin{equation}
\bar{\sigma}_{\rm acc} = \sqrt{\sum_{i=1}^{m} \Big( n_{c}(\delta t_{i}) 
- \bar{n}_{\rm acc} \Big ) ^2 / (m-1)} . \label{eq:dev}
\end{equation}

Since there is no real coincidence if $|\delta t|\gg\Delta t_{\rm dis}$,
the distribution of the number of coincidences with time delay can be considered 
as an estimation of the distribution of accidental coincidences. 
The number of coincident events, $n_c(0)$, 
is compared to the estimated distribution. 

\begin{figure}
\scalebox{0.5}[0.5]{\includegraphics{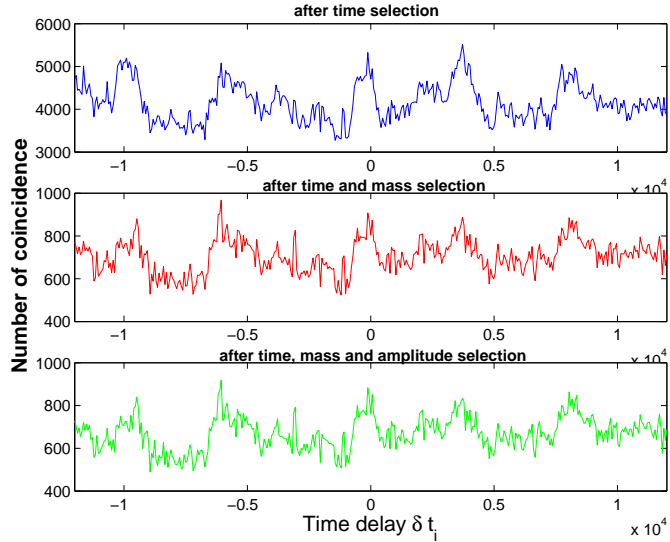}}
\caption{From top to bottom, the time delay histogram after time selection, 
after time and mass selection, and after time, mass and amplitude selection,
respectively, are plotted. }
\label{fig:timedelayhist25}
\end{figure}

\begin{figure}
\scalebox{0.5}[0.5]{\includegraphics{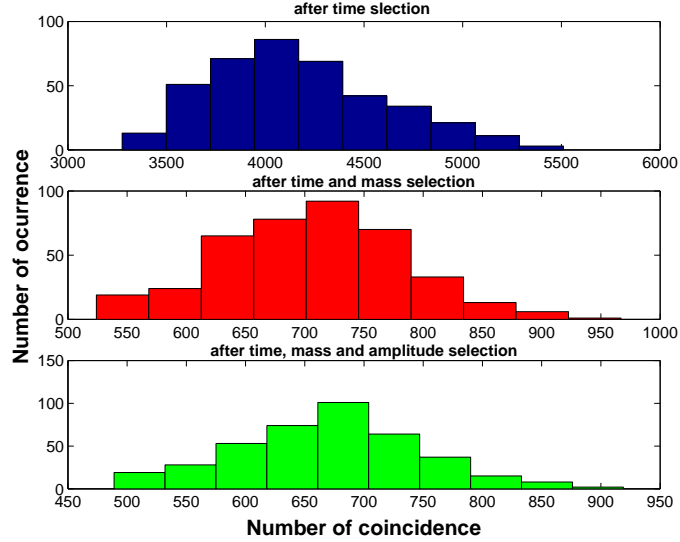}}
\caption{The distribution of the number of realizations
with 400 different time delays with respect to
the number of coincidences derived from Fig.~\ref{fig:timedelayhist25}.  
From top to bottom, the distribution
after time selection, after
time and mass selection, and and after time, mass and amplitude selection
are plotted.}
\label{fig:parentdist25}
\end{figure}

Fig. \ref{fig:timedelayhist25} shows the 
time delay histograms with $m=400$.
The 400 time delays are chosen from $-12000$ sec to $12000$ sec in 
increments of 60 seconds. 
The distribution of accidentals is shown in Fig.~\ref{fig:parentdist25}. 
In Table~\ref{tab:coincident-results}, we also list the expectation values of
 the number of accidental coincidence and the standard deviation
after each selection procedure. 
As can be seen from this, the number of coincident events 
after each selection procedures is consistent with the expected number of
accidental coincidences
within the statistical fluctuations.
Thus, we conclude that no statistically significant signals
of real coincident events are observed in our search. 

\section{Upper limit to the event rate from coincidence analysis}\label{sec:limit}

In this section, we present a method to evaluate the upper limit 
to the event rate based on the above result of the coincidence analysis. 

The upper limit to the event rate is given by Eq.~(\ref{eq:eventrate}) as in 
the case of the single-detector searches. 
The upper limit $N$ to the average number of real events
can be determined by Eq.~(\ref{eq:upperlimitformula}), using 
the observed number of events $N_{\rm obs}$ with $\rho/\sqrt{\chi^2}$ greater than the threshold, 
the estimated number of fake events $N_{\rm bg}$ with $\rho/\sqrt{\chi^2}$ greater than the threshold,
and the confidence level. 
We set different thresholds to the value of 
$\rho_{\mbox{\rm\tiny TAMA}}/\sqrt{\chi_{\mbox{\rm\tiny TAMA}}^2}$ and 
$\rho_{\mbox{\rm\tiny LISM}}/\sqrt{\chi_{\mbox{\rm\tiny LISM}}^2}$ respectively.
An advantage of this is that, because of its simplicity,
it can be readily applied to the cases when more than two detectors
with different arm directions are involved. 

We determine a background distribution $f(y_1,y_2)$ of the number of 
coincident events 
from the data for $y_1>5.5$ or $y_2>5.5$ in Fig.~\ref{fig:determthre25}, 
where $y_1=\rho_{\mbox{\rm\tiny TAMA}}/\sqrt{\chi_{\mbox{\rm\tiny TAMA}}^2}$
and $y_2=\rho_{\mbox{\rm\tiny LISM}}/\sqrt{\chi_{\mbox{\rm\tiny LISM}}^2}$. 
We evaluate the expected number of fake events which $\rho/\sqrt{\chi^2}$ is greater than the thresholds
$y_1=y_{\mbox{\rm\tiny T}}$ or $y_2=y_{\mbox{\rm\tiny L}}$ by
\begin{equation}
N_{bg}=
\int_{y_{\rm\scriptscriptstyle T}}^\infty dy_1\int_0^\infty dy_2 f(y_1,y_2)+
\int_{0}^\infty dy_1\int_{y_{\rm\scriptscriptstyle L}}^\infty dy_2 f(y_1,y_2)-
\int_{y_{\rm\scriptscriptstyle T}}^\infty dy_1 
\int_{y_{\rm\scriptscriptstyle L}}^\infty dy_2 f(y_1,y_2).
\end{equation}
As the false alarm rate, we adopt $0.00023$ [1/hour] ($=2.0$ [1/yr]) 
which corresponds to the number of 
expected fake events $N_{\rm bg}=0.063$. We choose the 
thresholds $y_1=y_{\mbox{\rm\tiny T}}=8.3$
for TAMA300 and 
$y_2=y_{\mbox{\rm\tiny L}}=8.1$ for LISM.
The observed number of events with $y_1$ or $y_2$ greater than the threshold is $N_{\rm obs}=0$. 
Therefore we obtain the upper limit to the 
average number of real events  with $y_1$ or $y_2$ greater than the threshold
 as $N=2.30\ (CL.=90\%)$
from Eq.~(\ref{eq:upperlimitformula}). 

\begin{figure}
\scalebox{0.5}[0.5]{\includegraphics{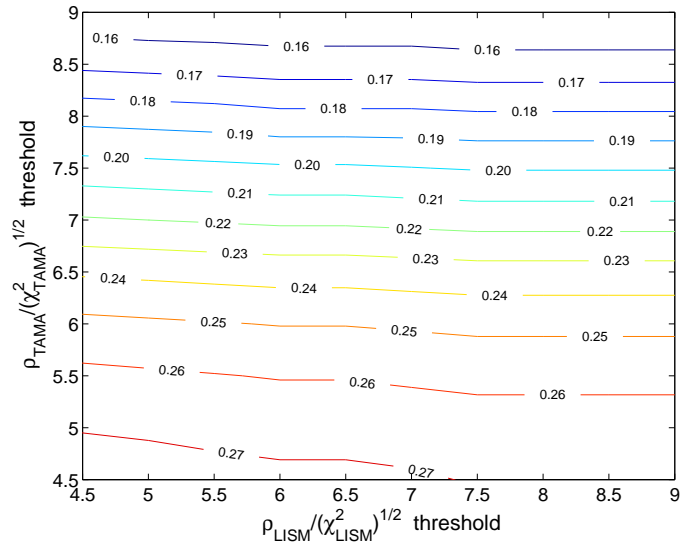}}
\caption{Detection efficiency in the coincidence analysis of
sources within 1kpc as a function of TAMA300 and LISM thresholds 
for $\rho/\sqrt{\chi^2}$.}
\label{fig:efficcoin}
\end{figure}

The detection probability $\epsilon$ is derived by the method 
explained in Section~\ref{sec:efficiency}, and is 
shown in Fig.~\ref{fig:efficcoin}. 
With the thresholds chosen above, 
we obtain $\epsilon= 0.182$.
Using the upper limit to the 
average number of real events $N$ with $y_1$ or $y_2$ greater than the threshold,
the detection probability $\epsilon$
and the length of data $T=275$ [hours],
we obtain an upper limit to the 
event rate within 1kpc to be 
$N/(T \epsilon) = 0.046$ [1/hour] (CL.= $90\%$).

Unfortunately, 
this value is not improved from the value obtained by 
the analysis of the TAMA300 data. 
The dominant effect that causes the difference in the upper limit
for a single-detector analysis and the coincidence analysis comes from 
the difference in the detection efficiency. 
The detection efficiency of the coincidence analysis in our case is 
determined by that of LISM, since LISM has the lower sensitivity. 
The efficiency of LISM is improved in the case of the coincidence analysis,
since the threshold is lowered.
However, this does not compensate the difference
in the detection efficiency between TAMA300 and LISM. 
The efficiency of TAMA300 
is already nearly 100 \% in 1kpc without performing the
coincidence analysis. 
Thus, by taking the coincidence with 
the detector which has much lower sensitivity, 
the detection efficiency of the coincidence analysis 
becomes lower than the case of TAMA300 alone. 
As a result, the upper limit to the event rate we obtained
by the coincidence analysis
is less stringent than the one obtained by the analysis of
the TAMA300 data. 

\section{Summary and discussion}
\label{sec:summary}

In this paper, we performed a coincidence analysis using
 the data of TAMA300 and
LISM taken during DT6 observation in 2001.

We analyzed the data from each detector by matched filtering
and obtained event lists. 
Each event in the lists was characterized by the time of coalescence, 
masses of the two stars, and the amplitude of events. 
If any of the events are true gravitational wave events, they should have 
the consistent values of these parameters in the both event lists. 
We proposed a method to set coincidence conditions for the 
source parameters such like the time of coalescence, chirp mass, 
reduced mass, and the amplitude of events. 
We took account of 
the time delay due to the distance between the two detectors,
the finite mesh size of the mass parameter space, 
the difference in the signal amplitudes
due to the different sensitivities 
and antenna patterns of the detectors,
and errors in the estimated parameters due to the instrumental noise.
Our Monte Carlo studies showed that we would 
not lose events significantly by imposing the coincidence conditions. 

By applying the above method of the coincidence analysis 
to the event lists of TAMA300 and LISM, 
we can reduce the number of fake events by a factor $10^{-4}$ compared with 
the number of fake events before the coincidence analysis.
In order to estimate the number of accidental coincidences 
produced by noise, we used the time shift procedure. 
We found that the number of events survived after imposing 
the coincidence conditions is consistent 
with the expected number of accidental coincidences
within the statistical fluctuations.
Thus we found no evidence of gravitational wave signals. 
As discussed in Appendix~\ref{sec:sidereal}, 
the sidereal time distribution of the survived events were also consistent 
with the distribution of accidentals. 

Finally, we proposed 
a simple method to set an upper limit to the event rate 
and applied it to the above results of the coincidence analysis.
We obtained an upper limit 
to the Galactic event rate within 1kpc from the Earth
to be 0.046 [1/hour] (90\%\,CL). 
In our case, since  LISM has a much lower sensitivity
than TAMA300, we were unable to obtain a more stringent upper limit
to the event rate than the one obtained by the single-detector analysis 
of TAMA300.
This is because the detection efficiency in the coincidence analysis
is determined by the detector with a lower sensitivity. 

However, if we have two detectors that have comparable sensitivities,
it is possible to obtain an improved upper limit
 compared to a single-detector analysis.
As an example, let us imagine the case
when the sensitivity of LISM is the same as that of TAMA300. 
The result of Galactic event simulations suggests that the
detection efficiency in the case of a single-detector analysis
is 0.35, while it improves to 0.48 in the case of a coincidence analysis. 
These values are translated to upper limits on the Galactic event rate
of 0.026 [1/hour] ($90\%$\,CL) for the single-detector case and
0.019 [1/hour] ($90\%$\,CL) for the two-detector case. 

The method of a coincidence analysis and the method to set an upper limit
to the event rate proposed here can be readily applied to
the case when there are more than two detectors with arbitrary arm 
directions. Hence these methods will be useful for data analysis
for a network of interferometeric gravitational wave detectors in the 
near future. 

\acknowledgments

This work was supported in part by the Grant-in-Aid for 
Scientific Research on Priority Areas (415) of the Ministry of Education, 
Culture, Sports, Science and Technology of Japan,
and in part by JSPS Grant-in-Aid
for Scientific Research Nos.~14047214, 12640269 and 11304013. 

\newpage
\appendix

\section{method to distinguish between real events and non-Gaussian noise}
\label{sec:Ap1}

The real data from TAMA300 and LISM 
contain non-stationary and non-Gaussian noise. 
One way to remove the influence of such noise is a
veto analysis by using the data of various channels which monitor 
the status of the interferometers and their environments. 
Such an analysis has been performed using the data of 
TAMA300 \cite{ref:vetoana}. 
However, more efforts will be needed
 to establish an efficient and faithful veto method. 

It was shown that about $20\,\%$ of the data from TAMA300 DT6 
contains non-Gaussian noise significantly~\cite{ref:andoburst}. 
Even if we remove this portion of the data with large
non-Gaussian noise, the rest of data may still 
contain some non-Gaussian noise.
It is thus necessary to introduce a method by which 
we can discriminate the non-Gaussian noise from real gravitational
wave signals using the properties of inspiral signals.
As one of such methods, the $\chi^2$ method was introduced 
in~\cite{ref:40m}. 

In this method, we examine whether 
the time-frequency behavior of the data is consistent with 
the expected signal. 
We divide each template into $n$ mutually independent pieces
in the frequency domain, 
chosen so that the expected contribution to $\rho$ from 
each frequency band is equal: 
\begin{equation}
\tilde{{h}}_{(c,s)}(f)=\tilde{{h}}^{(1)}_{(c,s)}(f)
+\tilde{{h}}^{(2)}_{(c,s)}(f)+
             \cdots +\tilde{{h}}^{(n)}_{(c,s)}(f). 
\end{equation}
We introduce 
\begin{equation}
z^{(i)}_{(c,s)}=(s,{h}^{(i)}_{(c,s)}), 
\quad 
\overline{z}^{(i)}_{(c,s)}={1\over n}(s,{h}_{(c,s)}). 
\end{equation}
Then, $\chi^2$ is defined by 
\begin{equation}
\chi^2=\sum_{i=1}^n \left[
{\left(z^{(i)}_{(c)}- \overline{z}^{(i)}_{(c)}\right)^2 + 
\left(z^{(i)}_{(s)}- \overline{z}^{(i)}_{(s)}\right)^2
\over \sigma_{(i)}^2}\right],
\end{equation}
with
\begin{equation}
\sigma_{(i)}^2=({h}^{(i)}_{(c)},{h}^{(i)}_{(c)})
=({h}^{(i)}_{(s)},{h}^{(i)}_{(s)})={1\over n}.
\end{equation}
Provided that the noise is Gaussian,
this quantity must satisfy the $\chi^2$-statistics with 
$2n-2$ degrees of freedom and is independent 
of $\rho=\sqrt{z_{(c)}^2+z_{(s)}^2}$.
For convenience, we use a reduced chi-square defined by 
$\chi^2/(2n-2)$. 
In this paper, we choose $n=16$. 

In the case of TAMA300, it was found that there was a 
strong tendency that noise events with large $\chi^2$ have
large values of $\rho$. 
Since the value of $\chi^2$
will be independent of the amplitude of inspiral signals 
when the parameters such as 
$t_c$, $M$ and $\eta$ of the signal are equal to those of 
a template~\cite{ref:grasp},
one may expect that we can discriminate 
real signals from noise events by rejecting events with 
large $\chi^2$, and this method was used in the TAMA300 DT2 
analysis~\cite{ref:DT2}.

However, in reality, since we perform analysis on a discrete 
$t_c$ and a discrete mass parameter space, 
the parameters of a signal do not coincide with
those of a template in general. 
We have found in the analysis of the TAMA300 DT4 data in 2000 
that this difference produces a large value of 
$\chi^2$ when the $SNR$ of an event is very large
even if the event is real~\cite{ref:TAMAinspiral}. 
Thus, if we apply a threshold to the value of $\chi^2$
to reject noise events, 
we may lose real events with large $SNR$. 
This is a serious problem since an event with a large $SNR$ 
has a high statistical significance of it to be real.
This lead us to introduce a different rejection criterion
when we performed an inspiraling wave search
with the TAMA300 DT4 data \cite{ref:TAMAinspiral}, namely,
 a threshold on the value of $\rho/\sqrt{\chi^2}$.
By Galactic event simulations, 
we found that this new criterion can give 
a better detection efficiency
of the Galactic events without losing strong amplitude events.

\begin{figure}
\scalebox{0.5}[0.5]{\includegraphics{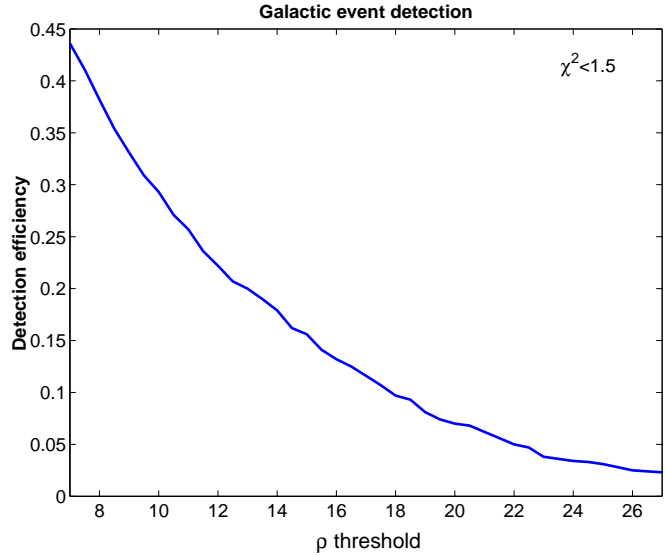}}
\caption{Detection efficiency for a $\chi^2$ threshold.
In this figure, the threshold is set to $\chi^2=1.5$.}
\label{fig:efficchi}
\end{figure}

Here we examine whether the $\rho/\sqrt{\chi^2}$ selection is 
useful also in the case of the TAMA300 DT6 data. 
For comparison, the detection efficiency for a
simple $\chi^2$ threshold is shown in Fig.~\ref{fig:efficchi}. 

For the $\chi^2<1.5$ threshold,
using 287.6 hours of the data,
the false alarm rate $2.0$ [1/yr] determines the
$SNR$ threshold to be $\rho = 12.5$. 
This gives the detection efficiency of $0.213$. 
On the other hand, as discussed in Section~\ref{sec:matchedfilterresults}, 
the detection efficiency in the case of the $\rho/\sqrt{\chi^2}$ threshold
is $0.263$ for the same false alarm rate, $2.0$ [1/yr]. 
We thus find that we have a better efficiency 
for the $\rho/\sqrt{\chi^2}$ threshold, 
although the gain of efficiency is not very large. 
However, the important point is that we have
much larger detection efficiency for signals with 
large $SNR$.

\section{Different choice of $\Delta t$}
\label{sec:3.2sec}

In this appendix,
we consider the case of a different choice
of the length of duration $\Delta t$
to find local maximum of matched filtering output (see \ref{sec:matchedfiltering}),
to see if our conclusion is affected by a different choice of $\Delta t$.

\begin{figure}
\scalebox{0.5}[0.5]{\includegraphics{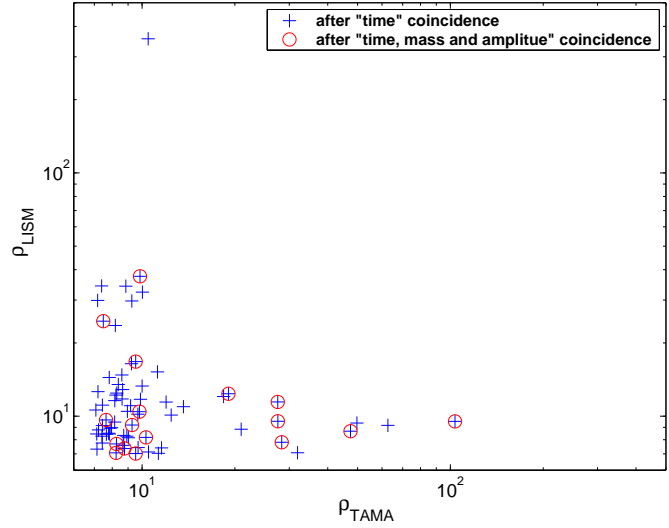}}
\caption{($\rho_{\mbox{\rm\tiny TAMA}}$,$\rho_{\mbox{\rm\tiny LISM}}$)
scatter plots in the case $\Delta t = 3.28$sec.
The crosses (+)
are the events survived after the time selection, 
and the circled crosses ($\protect\oplus$)
 are the events survived after the time, mass and amplitude 
selections.}  
\label{fig:scatterrhocoin3.2}
\end{figure}

\begin{figure}
\scalebox{0.5}[0.5]{\includegraphics{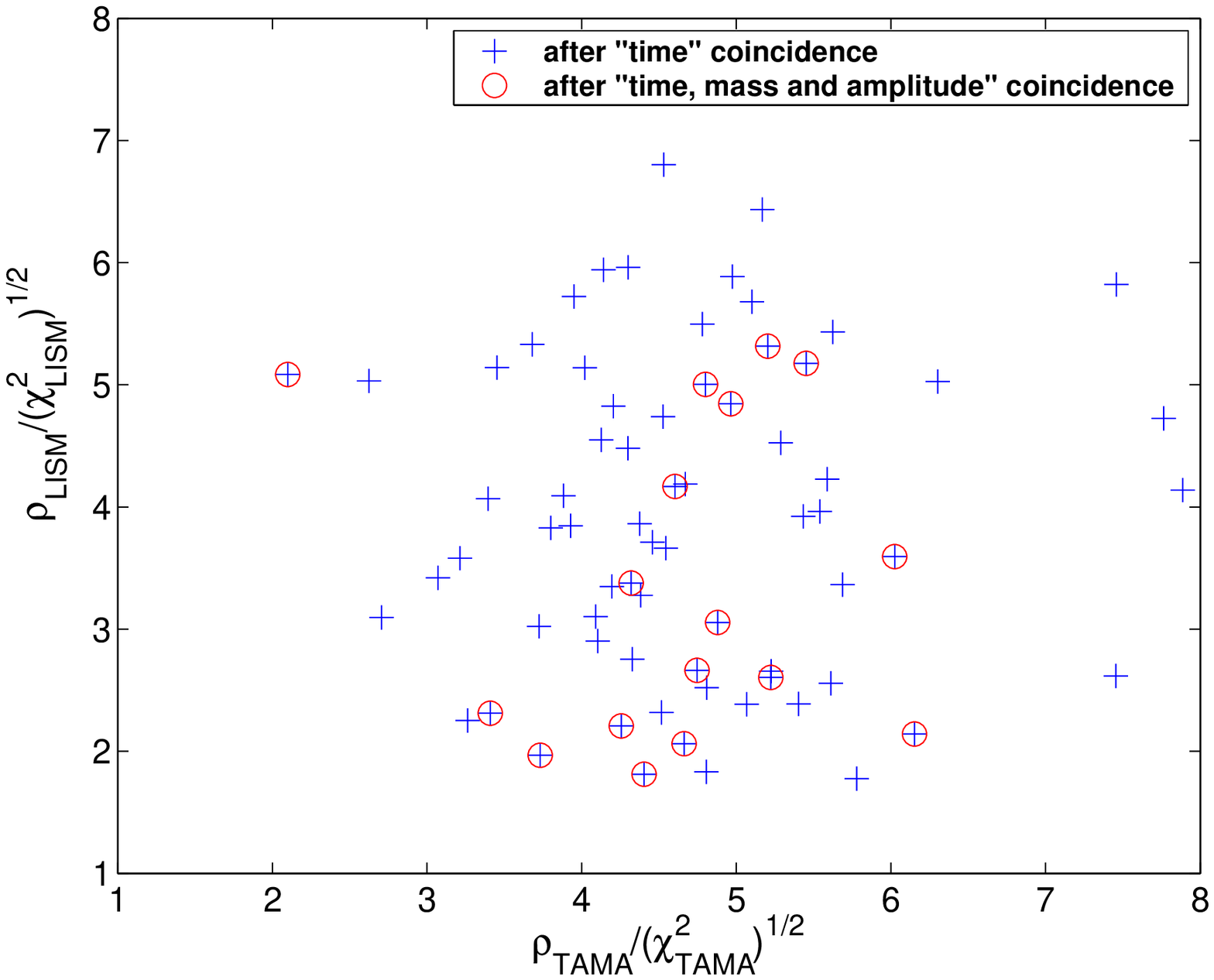}}
\caption{($\rho_{\mbox{\rm\tiny TAMA}}/\sqrt{\chi^2_{\mbox{\rm\tiny TAMA}}}$ 
,$\rho_{\mbox{\rm\tiny LISM}}/\sqrt{\chi^2_{\mbox{\rm\tiny LISM}}}$)
scatter plots in the case  $\Delta t = 3.28$sec,
The crosses (+)
are the events survived after the time selection, 
and the circled crosses ($\protect\oplus$)
 are the events survived after the time, mass and amplitude 
selections.}
\label{fig:determthre3.2}
\end{figure}

Here we adopt $\Delta t = 3.28$sec. In this case, the total
number of events is found 
to be 158,437 for TAMA300 and 142,465 for LISM.
The numbers of events survived after each step of the
coincidence selections are given in Table~\ref{tab:coincident-results2}. 
The corresponding estimated numbers of accidentals are also shown. 
The scatter plots of these selected events are shown
in Figs.~\ref{fig:scatterrhocoin3.2} and \ref{fig:determthre3.2}. 
We see that the number of coincident events is consistent with 
the number of accidentals within the standard deviation,
in agreement with our conclusion given in the main text of this paper.


\begin{table}[htpd]
\begin{center}
\renewcommand{\arraystretch}{1.5}
\begin{tabular}{c c c}
\multicolumn{3}{c}{Results of independent matched filtering searches}
\\
\hline \hline
 &TAMA300&LISM\\
\hline
Number of events & 158,437 & 142,465\\
\hline \hline
\\
\multicolumn{3}{c}{Results of coincidence analysis}
\\
\hline \hline
   & $n_{\rm obs}$ & $\bar{n}_{\rm acc} \pm \bar{\sigma}_{\rm acc}$    \\
\hline
after time selection & 70   & $75.0 \pm 8.6$ \\
after time and mass selection & 18  & $18.8 \pm 4.1$   \\
after time, mass and amplitude selection & 17 & $17.9  \pm 3.8$  \\
\hline \hline
\end{tabular} 
\end{center}
\caption{Results of coincidence analysis
in the case $\Delta t = 3.28$sec. 
$n_{\rm obs}$ is the number of coincidence events. 
$\bar{n}_{\rm acc}$, $\bar{\sigma}_{\rm acc}$ are the 
estimated number of accidental coincidence and its variance.}
\label{tab:coincident-results2} 
\end{table}



\section{Sidereal time distribution}
\label{sec:sidereal}

\begin{figure}
\scalebox{0.6}[0.6]{\includegraphics{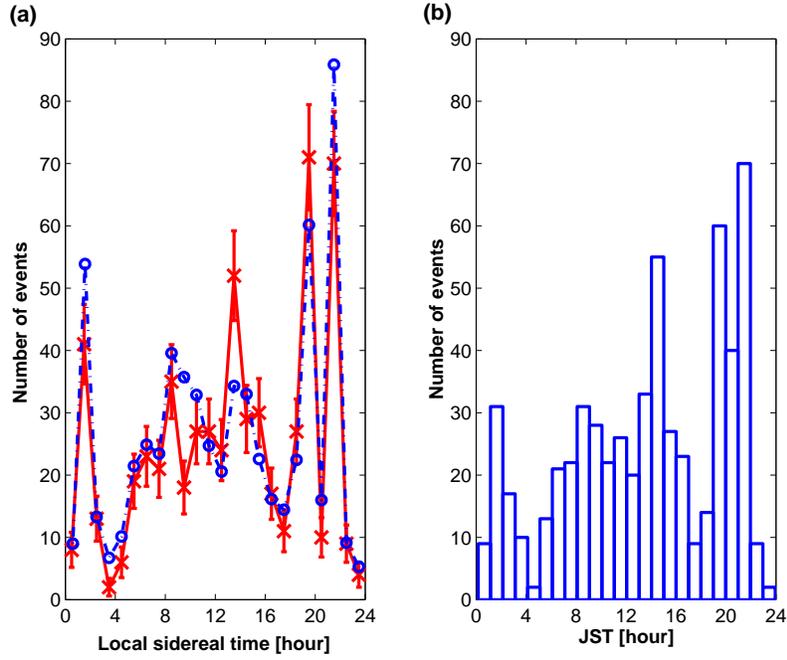}}
\caption{(a) The event distribution as a function of the 
local sidereal time. 
The solid line represents the number of coincident events per 
one sidereal hour. The dot-dashed line represents the 
estimated number of accidentals. 
(b) The number of coincident events as a function of 
the Japanese Standard Time.}
\label{fig:sidereal}
\end{figure}

In this appendix, we examine the sidereal time distribution
of the events. 
In Fig.~\ref{fig:sidereal}\,(a), 
we plot the number of coincident events 
as a function of the local sidereal hour
at the location of TAMA300.
The estimated number of accidental coincidences are also plotted,
which are obtained by the same time shift method used
 in Section~\ref{sec:coincidenceresults}
but for data within each bin of the sidereal hour. 
If the gravitational wave sources are sharply concentrated
in the Galactic disk, we would detect more events when
the zenith direction of the detector coincides with the direction to 
the Galactic plane than the rest of time.
The zenith direction faces to the Galactic disk at around 6:00 and 18:00
in the sidereal hour. 
Since LISM is only sensitive to sources within a few kpc, 
we may not be able to see any significant excess of the events in
the Galactic disk within this distance unless the concentration of
the sources to the Galactic disk is very strong. 
Even in this case, it is useful to investigate the sidereal time distribution
to look for signatures of real events. 

We find that the distribution of coincident events
is consistent with accidentals, 
although there are a few hours in which 
the agreement is not very good.
Thus, we conclude that the result of the sidereal hour distribution 
is consistent with the number of accidentals, and there is
no signature of gravitational wave event.

In Fig. \ref{fig:sidereal}\,(b),
we also plot the number of coincident events as a function of the 
Japanese Standard Time (JST). 
Since the deviation of the local sidereal time 
from JST is not very large 
during the period of observation,
this figure is very similar to Fig.~\ref{fig:sidereal}\,(a). 
The reason that there are many coincident events during 20:00 to 22:00
JST is due to a large number of events recorded by LISM during that period.
During the DT6 observation, there were some activities in the Kamioka mine
from 20:00 to 22:00 JST, and trucks went through the tunnel of the mine 
during that period. We suspect this caused fake events in LISM. 

\section{Parameter estimation errors induced by detector noise}
\label{sec:Ap2}

In this appendix, we briefly review the theory of the parameter
 estimation error developed in~\cite{ref:Cutler}. 
This is used in determining the parameter windows for
 the coincidence analysis in this paper. 

In the matched filtering, 
for a given incident gravitational wave, different realizations of the noise
will give rise to somewhat different best-fit parameters.
For a large $SNR$, the best-fit parameters will have Gaussian 
distributions centered on the correct values. Specifically, let 
$\tilde{\theta^i}$ be the correct values of the parameters,
and let $\tilde{\theta^i} + \Delta \theta^i$ be
the best-fit parameters in the presence of a realization of noise.
Then for large $SNR$, the parameter estimation errors 
$\Delta \theta^i$ have the Gaussian probability distribution 
\begin{equation}
p(\Delta \theta^i) = \mathcal{N}e^{-\frac{1}{2}
 \Gamma_{ij} {\Delta \theta^i\Delta \theta^j}} .
\end{equation}
where $\Gamma_{ij}$ is the called Fisher Information matrix defined by
\begin{equation}
\Gamma_{ij} \equiv \Big( \frac{\partial h}{\partial \theta^i},
 \frac{\partial h}{\partial \theta^j} \Big) ,
 \label{eq:Fisher}
\end{equation} 
and $\mathcal{N} = \sqrt{{\rm det}(\mathbf{\Gamma}/2\pi)}$ is the
normalization factor. It follows that the root-mean-square errors in 
$\theta^i$ is given by
\begin{equation}
\sigma_{i} 
= \sqrt{((\Delta \theta^i)^2)} = \sqrt{\Sigma^{ii} }
\label{eq:error} ,
\end{equation} 
where $\mathbf{\Sigma} \equiv \mathbf{\Gamma}^{-1}$, and 
the correlation coefficient between parameters $\theta^i$ and $\theta^j$ is 
given by
\begin{equation}
c^{ij}=\frac{<\Delta \theta ^i \Delta \theta ^j>}{\sigma_i \sigma_j}
=\frac{\Sigma^{ij}}{\sqrt{\Sigma^{ii} \Sigma^{jj}}}.  \label{eq:corr}
\end{equation}
By definition, each $c^{ij}$ lies in the range $(-1,1)$.

As given in Section~\ref{sec:basicformula},
an inspiraling signal in the frequency domain is given by 
\begin{equation}
\tilde{h}(f) = \mathcal{A} f ^{-7/6} e ^{i \psi(f)} . \label{eq:hf}
\end{equation}
Here we consider the phase $\psi(f)$ up only to the
second post-Newtonian order but including the effect of the spins
of stars.
Note that this is slightly different from the template formula
(\ref{eq:2.5PN}) used in our analysis.
The phase $\psi(f)$ is given by 
\begin{eqnarray}
\psi (f) &=& 2\pi ft_c - \phi_c - \frac{\pi}{4} 
+ \frac{3}{128} (\pi \mathcal{M} f) ^{-5/3} 
\Big[ 1+ \frac{20}{9} \Big( \frac{743}{336} + \frac{11}{4}\eta \Big)
 (\pi M f)^{2/3} - 4(4 \pi - \beta ) (\pi M f)
 \nonumber \\
        & &{}+10 \Big(\frac{3058673}{1016064} + \frac{5429}{1008} \eta 
+ \frac{617}{144} \eta^2 - \sigma \Big)(\pi M f)^{4/3} \Big]. 
\end{eqnarray}
In the above, $\beta$ is the spin-orbit parameter given by 
\begin{equation}
\beta = \frac{1}{12}\sum _{i=1}^2 [113 (m_i/M) + 75 \eta ] 
\hat{\mathbf{L}} \cdot \Vec{\chi} _i , \label{eq:spin-orbit}
\end{equation}
and $\Vec{\chi} _i = \mathbf{S} _i /m^2_i$, and $\mathbf{S}_i$ is the 
spin angular momentum of each star, and $\hat{\mathbf{L}}$ is 
the unit vector along the orbital angular momentum vector. 
The spin-spin parameter $\sigma$ is given by 
\begin{equation}
\sigma = \frac{\eta}{48} (-247\Vec{\chi} _1 \cdot \Vec{\chi} _2 
+ 721\hat{\mathbf{L}} \cdot \Vec{\chi} _1 
\hat{\mathbf{L}} \cdot \Vec{\chi} _2 ) . \label{eq:spin-spin} 
\end{equation}

We define 
\begin{equation} 
\rho ^2 = 4 \mathcal{A} ^2 \int^{f_{\rm max}}_{0} 
\frac{f^{-7/3}}{S_{n} (f)}df .
\end{equation}
We also define the frequency moments $\bar{f} _{\alpha}$ of the noise spectrum
density:
\begin{eqnarray}
f_{7/3} &\equiv & \int _{0} ^{f_{\rm max}} df 
\quad [f ^{7/3} S _{n} (f) ]^{-1} \\
\bar{f} _{\alpha} &\equiv & f_{7/3}^{-1} \int _{0} ^{f_{\rm max}} df 
\quad [f ^{\alpha} S _{n} (f) ]^{-1} .
\end{eqnarray}

In order to evaluate the Fisher matrix, 
we calculate the derivatives of $\tilde{h}(f)$ with respect to
the seven parameters 
\begin{equation}
\Vec{\theta} = (\ln{\mathcal A},f_0 t_c, \phi_c, \ln {\cal M},
\ln \eta , \beta , \sigma ),
\label{eq:parameter}
\end{equation}
where $f_0$ is a fiducial frequency which is 
taken to be the frequency at which $S_n (f)$ becomes minimum.
We obtain
\begin{eqnarray}
\frac{\partial \tilde{h}(f)}{\partial \ln \mathcal{A}} &=& \tilde{h} (f) ,
 \nonumber \\
\frac{\partial \tilde{h}(f)}{\partial f_0 t_c}
 &=& 2\pi i \Big(\frac{f}{f_0}\Big)\ \tilde{h} (t) ,
 \nonumber \\
\frac{\partial \tilde{h}(f)}{\partial \phi _c} &=& - i \tilde{h} (f) ,
 \nonumber \\
\frac{\partial \tilde{h}(f)}{\partial \ln \mathcal{M}} 
&=& - \frac{5 i}{128} (\pi \mathcal{M} f)^{-5/3}
\big\{1+A_4(\pi M f)^{2/3} - B_4(\pi M f) +C_4 (\pi M f)^{4/3} \big\}\ 
 \tilde{h} (f) , 
\nonumber \\
\frac{\partial \tilde{h}(f)}{\partial \ln \eta} 
&=& - \frac{i}{96} (\pi \mathcal{M} f)^{-5/3}
\big\{A_5(\pi M f)^{2/3} - B_5(\pi M f) +C_5 (\pi M f)^{4/3} \big\}\ 
\tilde{h} (f) ,
 \nonumber \\
\frac{\partial \tilde{h}(f)}{\partial \beta} 
&=& \frac{3 i}{32} \eta ^{-3/5} (\pi \mathcal{M} f)^{-2/3} \tilde{h} (f) ,
 \nonumber \\
\frac{\partial \tilde{h}(f)}{\partial \sigma}
 &=& - \frac{15 i}{64} \eta ^{-4/5} (\pi \mathcal{M} f)^{-1/3} \tilde{h} (f) .
\end{eqnarray}
Here we have defined
\begin{eqnarray}
A_4 &=& \frac{4}{3} \Big(\frac{743}{336} + \frac{11}{4} \eta \Big) ,
 \nonumber \\
B_4 &=& \frac{8}{5} (4\pi - \beta), 
 \nonumber \\
C_4 &=& 2 \Big( \frac{3058673}{1016064} 
+ \frac{5429}{1008} \eta + \frac{617}{144} \eta ^2 - \sigma \Big) ,
\end{eqnarray}
and
\begin{eqnarray}
A_5 &=& \frac{743}{168} - \frac{33}{4} \eta ,
 \nonumber \\
B_5 &=& \frac{27}{5}(4\pi - \beta),
  \nonumber \\
C_5 &=& 18  \Big( \frac{3058673}{1016064}
 + \frac{5429}{4032} \eta + \frac{617}{96} \eta ^2 - \sigma \Big) .
\end{eqnarray}

Finally, the components of $\Vec{\Gamma}$ can be obtained by
evaluating Eq. (\ref{eq:Fisher}).
They can be expressed
in terms of the parameters $\Vec{\theta}$, the signal-to-noise
ratio $\rho$, and the frequency moments $\bar{f} _{\beta}$.
The components of $\Gamma_{ij}$ are given by 
\begin{eqnarray}
\Gamma_{ \ln \mathcal{A}j} &=& \delta _{\ln \mathcal{A}j} \rho ^2 ,
\qquad (j = \ln \mathcal{A}, f_0 t_c ,\phi_c , \ln \mathcal{M} ,
\ln \eta ,\sigma ,\beta ) , 
\label{eq:gammai}
\end{eqnarray}
\begin{eqnarray}
\Gamma_{t_c t_c } &=& (2 \pi )^2 \frac{1}{f_0} \bar{f} _{1/3} \ \rho ^2 , 
\\
\Gamma_{t_c \phi_c} &=& -2 \pi \frac{1}{f _0} \bar{f} _{4/3} \ \rho ^2 , 
\\
\Gamma_{ t_c \ln \mathcal{M}} &=& - \frac{5 \pi }{64 f_0}
 (\pi \mathcal{M} )^{-5/3} \Big( \bar{f} _3 +A_4 (\pi M) ^{2/3}
 - B_4 (\pi M) \bar{f} _2 +C_4 (\pi M) ^{4/3} \bar{f}_{5/3} \Big) \ \rho ^2 ,
 \\
\Gamma_{t_c \ln \eta} &=& -\frac{\pi}{48 f _0} (\pi \mathcal{M} ) ^{-5/3}
 \Big(A_5 (\pi M ) ^{2/3} -B_5 (\pi M ) \bar{f} _2 + C_5 ( \pi M ) ^{4/3}
 \bar{f}_{5/3} \Big) \ \rho ^2 ,
 \\
\Gamma_{t_c \ln \beta} &=& \frac{3 \pi}{16 f_0} \eta ^{-3/5}
 (\pi \mathcal{M} ) ^{-2/3} \bar{f} _2 \ \rho ^2 ,
 \\
\Gamma_{ t_c \sigma} &=& - \frac{15 \pi}{32 f_0} \eta ^{-4/5}
 (\pi \mathcal{M} ) ^{-1/3} \bar{f} _{5/3} \ \rho ^2 ,
\end{eqnarray}
\begin{eqnarray}
\Gamma_{\phi_c t_c} &=& \Gamma _{t_c \phi _c} ,
 \\
\Gamma_{\phi_c \phi_c} &=& \rho ^2 ,
\\
\Gamma_{\phi_c \ln \mathcal{M} } &=& \frac{5}{128} (\pi \mathcal{M} )^{-5/3}
 \Big( \bar{f} _4 +A_4 (\pi M) ^{2/3} \bar{f} _{10/3}
 - B_4 (\pi M) \bar{f} _3 +C_4 (\pi M) ^{4/3} \bar{f}_{8/3} \Big) \ \rho ^2 ,
 \\
\Gamma _{\phi_c \ln \eta} &=& \frac{1}{96} (\pi \mathcal{M} ) ^{-5/3}
 \Big(A_5 (\pi M ) ^{2/3} \bar{f} _{10/3} -B_5 (\pi M ) \bar{f} _3
 + C_5 (\pi M ) ^{4/3} \bar{f} _{8/3} \Big) \ \rho ^2 ,
\\ 
\Gamma _{\phi_c \beta} &=& -\frac{3}{32} \eta ^{-3/5}
 (\pi \mathcal{M} ) ^{-2/3} \bar{f} _3 \ \rho ^2  ,
\\
\Gamma _{ \phi_c \sigma} &=& \frac{15}{64} \eta ^{-4/5}
 (\pi \mathcal{M} ) ^{-1/3} \bar{f} _{8/3} \ \rho ^2 ,
\end{eqnarray}
\begin{eqnarray}
\Gamma_{\ln \mathcal{M} t _c} &=& \Gamma _{t_c \ln \mathcal{M}} ,
 \\
\Gamma_{\ln \mathcal{M} \phi _c} &=& \Gamma _{\phi _c \ln \mathcal{M}},
\\
\Gamma_{\ln \mathcal{M} \ln \mathcal{M}} &=&
 \frac{25}{16384 (\pi \mathcal{M} )^{10/3}}
 \Big( A_4 ^2 (\pi M) ^{4/3} \bar{f} _{17/3}
 \bar{f} _{13/3} +2 A_4 (\pi M) ^{2/3} \Big\{\bar{f} _{15/3}
 - B_4 \pi M \bar{f} _4
 \nonumber  \\
& &{} +C_4 (\pi M) ^{4/3} \bar{f} _{11/3} 
+ \pi M (B_4 ^2 \pi M \bar{f} _{11/3} - 2 B_4 ( \bar{f} _{14/3}
 \nonumber \\
& &{}+ C_4 (\pi M) ^{4/3} \bar{f} _{10/3}))\Big\}\Big) \ \rho ^2 ,
\\
\Gamma_{\ln \mathcal{M} \ln \eta} &=&
  \frac{1}{12288 \pi ^{8/3} \mathcal{M} ^{10/3}}
 \Big( 5 M^{2/3}(A_5 \Big\{\bar{f} _{15/3}
 + A_4 (\pi M)^{2/3} \bar{f} _{13/3} - B_4  \pi M \bar{f} _{4}
  \nonumber  \\
& &{}+ C_4 (\pi M)^{4/3}  \bar{f} _{11/3} + (\pi M)^{1/3}
 (C_5 (\pi M)^{1/3}( \bar{f} _{13/3} + A_4 (\pi M)^{2/3} \bar{f} _{11/3}
 \nonumber \\
& &{} - B_4 \pi M \bar{f} _{10/3}+ C_4 (\pi M)^{4/3} \bar{f} _{3})
 - B_5 (\bar{f} _{114/3} + A_4(\pi M)^{2/3} \bar{f} _{4}
 \nonumber  \\
& &{} - B_4 \pi M \bar{f} _{11/3}
 +C_4 (\pi M)^{4/2} \bar{f} _{10/3})))\Big\}\Big) \ \rho ^2 ,
\\
\Gamma_{\ln \mathcal{M} \beta} &=& - \frac{15}{4096} \eta ^{-3/5}
 (\pi \mathcal{M} )^{-7/3}
 \Big( \bar{f} _{14/3} +A_4 (\pi M) ^{2/3} \bar{f} _{4}
 - B_4 (\pi M) \bar{f} _{11/3}
 \nonumber  \\
& &{}+C_4 (\pi M) ^{4/3} \bar{f}_{10/3} \Big) \ \rho ^2 ,
\\
\Gamma_{\ln \mathcal{M} \sigma} &=& \frac{75}{8192} \eta ^{-4/5}
 (\pi \mathcal{M} )^{-2} \Big( \bar{f} _{13/3}
 +A_4 (\pi M) ^{2/3} \bar{f} _{11/3} - B_4 (\pi M) \bar{f} _{10/3}
 \nonumber \\
& &{} +C_4 (\pi M) ^{4/3} \bar{f}_{3} \Big) \ \rho ^2 ,
\end{eqnarray}
\begin{eqnarray}
\Gamma_{\ln \eta t_c} &=& \Gamma _{t_c \ln \eta} ,
\\
\Gamma_{\ln \eta \phi _c} &=& \Gamma _{\phi _c \ln \eta} , 
\\
\Gamma_{\ln \eta \ln \mathcal{M}} &=& \Gamma _{\ln \mathcal{M} \ln \eta}  ,
\\
\Gamma_{\ln \eta \ln \eta} &=& \frac{1}{9216 \pi ^2 \mathcal{M} ^{10/3}}
 \Big(M ^{4/3} \Big\{ A_5^2 \bar{f} _{13/3} -2 A_5 B_5 (\pi M)^{1/3}
 \bar{f} _{4} +B_5^2 (\pi M)^{2/3} \bar{f} _{11/3}
 \nonumber \\
& &{}+ 2 A_5 B_5(\pi M)^{2/3} \bar{f} _{11/3}
 -2 B_5 C_5 \pi M  \bar{f} _{10/3}
 + C_5^2 (\pi M)^{4/3} \bar{f} _{3} \Big\} \Big) \ \rho ^2 , 
\\
\Gamma_{\ln \eta \beta} &=& -\frac{1}{1024} \eta ^{-3/5}
 (\pi \mathcal{M} ) ^{-7/3} \Big(A_5 (\pi M ) ^{2/3} \bar{f}_4
 -B_5 (\pi M ) \bar{f} _{11/3}
\nonumber \\
& &{} + C_5 ( \pi M ) ^{4/3} \bar{f} _{10/3} \Big) \ \rho ^2 , 
\\
\Gamma_{\ln \eta \sigma} &=& \frac{5}{2048} \eta ^{-4/5}
 (\pi \mathcal{M} ) ^{-2} \Big(A_5 (\pi M ) ^{2/3} \bar{f}_{11/3}
 -B_5 (\pi M ) \bar{f} _{10/3}
 \nonumber \\
& &{}+ C_5 ( \pi M ) ^{4/3} \bar{f} _{3} \Big) \ \rho ^2 ,
\end{eqnarray}
\begin{eqnarray}
\Gamma_{\beta t_c} &=& \Gamma _{t_c \beta}  ,
\\
\Gamma_{\beta \phi _c} &=& \Gamma _{\phi _c \beta},
 \\
\Gamma_{\beta \ln \mathcal{M}} &=& \Gamma _{\ln \mathcal{M} \beta} ,
 \\
\Gamma_{\beta \ln \eta} &=& \Gamma _{\ln \eta \beta} ,
 \\
\Gamma_{\beta \beta} &=& \frac{9}{1024} \eta ^{-6/5}
 (\pi \mathcal{M} ) ^{-4/3} \bar{f} _{11/3} \ \rho ^2 ,
\\
\Gamma_{\beta \sigma} &=& -\frac{45}{2048} \eta ^{-7/5}
 (\pi \mathcal{M} ) ^{-1} \bar{f} _{10/3} \ \rho ^2 ,
\end{eqnarray}
\begin{eqnarray}
\Gamma_{\sigma t_c} &=& \Gamma _{t_c \sigma} ,
\\
\Gamma_{\sigma \phi _c} &=& \Gamma _{\phi _c \sigma} ,
\\
\Gamma_{\sigma \ln \mathcal{M}} &=& \Gamma _{\ln \mathcal{M} \sigma} ,
\\
\Gamma_{\sigma \ln \eta} &=& \Gamma _{\ln \eta \sigma} ,
\\
\Gamma_{\sigma \beta} &=& \Gamma _{\beta \sigma} ,
\\
\Gamma_{\sigma \sigma} &=& \frac{225}{4096} \eta ^{-8/5}
 (\pi \mathcal{M} ) ^{-2/3} \bar{f} _{3} \ \rho ^2  .
\label{eq:gammaf}
\end{eqnarray} 
It is ensured by these formulas 
that the eigenvalues of the Fisher matrix are always
positive definite.

The variance-covariance matrix $\Sigma^{ij}$ can now be
obtained from $\Vec{\Sigma} = \Vec{\Gamma} ^{-1}$, and the 
root-mean square errors 
and the correlation coefficients are computed from Eqs.~(\ref{eq:error})
and (\ref{eq:corr}).

For example, using 
a typical noise spectrum density of TAMA300, 
the root-mean square errors of the parameters in the case $\rho=10$ and 
$\beta = \sigma = 0$ are evaluated to be
$\Delta {\cal A}^{\rm TAMA}/{\cal A}^{\rm TAMA}=0.10$, 
$\Delta t_{c}^{\rm TAMA}= 0.65$msec, 
$\Delta \phi_{c}^{\rm TAMA}= 6.88$radians, 
$\Delta\mathcal{M}^{\rm TAMA}/\mathcal{M}^{\rm TAMA}=1.43\times10^{-2}$, 
and $\Delta\eta^{\rm TAMA}/\eta^{\rm TAMA}=2.47\times10^{-1}$.

\end{document}